\documentclass{jpp}
\usepackage{amsfonts, amsmath, amssymb}
\usepackage{natbib, hyperref}
\usepackage{graphicx}
\usepackage{color}
\usepackage{bm}

\shorttitle{Abrupt Transitions}
\shortauthor{A. Plummer, J. B. Marston, S. M. Tobias}

\title{Joint Instability and Abrupt Nonlinear Transitions in a Differentially Rotating Plasma}

\author{A. Plummer\aff{1}, J.B. Marston\aff{2}, and S.M. Tobias\aff{3}}

\affiliation{\aff{1}Department of Physics, Harvard University, Cambridge, MA 02138, USA
\aff{2}Department of Physics, Brown University, Providence, RI 02912, USA 
\aff{3}Department of Applied Mathematics, University of Leeds, Leeds LS2 9JT, UK}

\begin{document}

\maketitle

\begin{abstract}
Global magnetohydrodynamic (MHD) instabilities are investigated in a computationally tractable two-dimensional model of the solar tachocline. The model's differential rotation yields stability in the absence of a magnetic field, but if a magnetic field is present, a joint instability is observed.  We analyze the nonlinear development of the instability via fully nonlinear direct numerical simulation, the generalized quasilinear approximation (GQL), and direct statistical simulation (DSS) based upon low-order expansion in equal-time cumulants. As the magnetic diffusivity is decreased, the nonlinear development of the instability becomes more complicated until eventually a set of parameters are identified that produce a previously unidentified long-term cycle in which energy is transformed from kinetic energy to magnetic energy and back. We find that the periodic transitions, which mimic some aspects of solar variability-- for example, the quasiperiodic seasonal exchange of energy between toroidal field and waves or eddies-- are unable to be reproduced when eddy-scattering processes are excluded from the model.\end{abstract}

\section{Introduction}
\label{Introduction}

It is well known in the field of plasma physics -- both laboratory and astrophysical -- that many magnetic configurations are susceptible to instability. Indeed, ideal instabilities driven by current or pressure gradients may provide an operational limit for the magnetic configurations in plasma devices. Challenges in fusion research include both identifying stable plasma configurations and controlling the nonlinear development of instabilities. 

In an astrophysical context, magnetic fields may also become dynamically unstable. One example is the loss of stability of magnetic fields in stellar atmospheres that may lead to huge releases of energy (and indeed flaring behavior) \citep[see e.g.][]{cowleyetal_2003}. The description of the loss of stability in the astrophysical context draws heavily on the pioneering theory from fusion plasmas such as \citet{fkr63, T68, TN2015}.

A second example is the instability of magnetic configurations in stellar radiative zones \citep{tay1973} that potentially leads to the generation of turbulence there and even dynamo action \citep{Dikpati_Gilman2001}. These instabilities are related to the current-driven instabilities of \citet{tay1973,tay1980,pt1985}. These authors examined the stability of toroidal magnetic fields to non-axisymmetric perturbations, both in cylindrical and spherical geometries (see also the extensive discussion in \citet{spruit1999}). Current-driven instabilities use the magnetic field as their energy source, with a strong magnetic field required for the instability to proceed;  the role of rotation is simply to mediate the rate at which energy can be extracted.

A related set of instabilities, which are most relevant to this report, have magnetic configurations that are stable in isolation but can be destabilized by the presence of a differential rotation
(see e.g. \citet{GF97,Gilman_2002,cally2003clamshell, cdg2008, hc2009}. These  {\it joint
instabilities} occur for relatively weak magnetic fields provided that the differential rotation is sufficiently strong. Here, the axisymmetric differential rotation and magnetic field, which in isolation are linearly stable, are together jointly unstable. The toroidal magnetic field therefore acts as a conduit to allow the extraction of energy from the differential rotation, though some energy may also be extracted from the current. In recent years, significant attention has focused on these instabilities, as it is believed that they may be important in the solar tachocline \citep{T2005}.

The tachocline is the thin layer of radial and latitudinal shear in the Sun at the base of the solar convection zone \citep{SZ1992,T2005}. Estimates place the tachocline at 0.693 $\pm$ 0.002 solar radii with a width of 0.039 $\pm$ 0.013 solar radii \citep{Charbonneau_1999}. Global magnetohydrodynamic (MHD) instabilities in this layer have been invoked in models to drive turbulence below the base of the convection zone and have also been used to provide an electromotive force that may contribute to the solar dynamo \citep{Dikpati_Gilman2001}. 

Investigations into these instabilities began following Watson's linear study on purely hydrodynamic instabilities in two-dimensional shells with differential rotation, as is found in the tachocline layer.  With a simple quadratic angular frequency, a two-dimensional shell was found to be hydrodynamically stable at parameter values appropriate to the tachocline \citep{Watson}. In a landmark paper, \citet{GF97} extended Watson's analysis by adding a toroidal field profile, stable in the absence of rotation, to the model. Their analysis determined that the MHD shell may be unstable to a non-axisymmetric joint instability if a large enough magnetic field is present. For plausible magnetic field configurations, even the differential rotation of the tachocline may provide a conduit to instability. It is also reasonable to expect that such instabilities arise in other solar-type stars with radiative interiors and convective outer envelopes. This instability can be found both for ideal MHD configurations as well as in models with diffusion, though in the latter case some forcing is required to maintain the basic state. 

We comment here on the physics of the non-axisymmetric joint instability. The presence of the magnetic field allows the instability to proceed, and enables the subsequent extraction of energy from the differential rotation. This is clearly reminiscent of the physics of the magnetorotational instability (MRI) \citep{velikhov59, chan60, bh91}, where coupling induced by even a weak magnetic field can trigger an instability in a hydrodynamically stable configuration. However, the instability we consider here is more akin to the MRI with a toroidal magnetic field as studied by \citet{op96}; in that case a non-axisymmetric instability arises that allows the system to access a turbulent lower energy state. As for that case, it is conjectured that, for the non-dissipative system, the joint instability acts for arbitrarily weak magnetic fields.

More recently, computational advances have made thorough nonlinear studies of MHD instabilities possible. Notably, \citet{Cally_2001} examined a variety of magnetic field profiles, identifying a clamshell instability in a two-dimensional model with a broad toroidal field. Using a three-dimensional thin shell model with an additional poloidal field, \citet{Miesch_2007} was able to reach a statistical steady state in which the clamshell instability could operate in a sustained manner.  In this state, he observed quasiperiodic solutions in which energy was exchanged between magnetic, kinetic and potential energies in the nonlinear regime. Similarly, a ``seasonal'' quasiperiodic exchange of energies resulting from the joint instability has recently been linked to solar observations by \citet{dikpati_etal_2017}.

In this paper, we describe striking and, to the best of our knowledge, novel behavior of the joint instability in a two-dimensional nonlinear model without an imposed poloidal field. We explore the emergence of a multidecadal cycle in which energy is transformed from kinetic energy to magnetic potential energy, with sharp transitions between two distinct modes.

We also use this joint instability as a paradigm problem to investigate the effectiveness of a range of methods in describing the statistics of the nonlinear saturation in a turbulent state. In addition to linear stability analysis and fully nonlinear direct numerical simulation, a number of other methods for understanding the behavior of a system of partial differential equations exist. In this paper, we will also apply generalized quasilinear approximations, as well as a class of methods known as direct statistical simulation (DSS). DSS can be complementary to direct numerical simulation, as it solves for the evolution of the statistics of the system, rather than integrating the equations of motion themselves forward in time. This provides a number of advantages, both in terms of computational efficiency and physical insight. \citet{tobias2011} and \citet{Constantinou:2018kh} fruitfully used DSS to study idealized, stochastically-driven, astrophysical flows, calling for a more thorough characterization of DSS methods in these systems. 

In the next section, we introduce the model of the two-dimensional instability. We proceed in Section \ref{Methods} to describe the range of tools we employ to investigate the nonlinear development of the instability. We describe the transitions that occur and how well these are captured in Section \ref{Results}, before drawing conclusions in Section \ref{Conclusions}.

\section{Description of the Model}
\label{Model}

We consider the simplest possible model of the joint instability of differential rotation and magnetic fields. The evolution of the system, in the absence of forcing, is described by the MHD equations
\begin{equation}
\frac{\partial \textbf{\textit{u}}}{\partial t}+(\textbf{\textit{u}} \cdot \nabla) \textbf{\textit{u}}  + 2\bm{\Omega} \times \textbf{\textit{u}}=-\nabla p+\textbf{\textit{j}} \times \textbf{\textit{B}}  + \nu \nabla^{2}\textbf{\textit{u}} ,
\end{equation}
\begin{equation}
\frac{\partial \textbf{\textit{B}}}{\partial t}=\nabla \times(\textbf{\textit{u}}\times\textbf{\textit{B}})+\eta \nabla^{2}\textbf{\textit{B}} .
\end{equation}
where \textbf{\textit{u}} is the velocity in the frame of the rotating shell,  $\bm{\Omega}$ is the angular rate of rotation of the shell, $\nu$ is viscosity, $\eta$ is magnetic diffusivity, \textbf{\textit{B}} is the magnetic field, and $\textbf{\textit{j}} = \mathbf{\nabla} \times \textbf{\textit{B}}$ is the current density. Here the density has been set to unity without loss of generality. 

We consider evolution on a two-dimensional spherical surface so that the momentum and induction equation 
can be described by the evolution of two scalar equations, one for the absolute vorticity $q(\theta,\phi,t)$ (with
$\theta$ being co-latitude and $\phi$ longitude), and one for the scalar magnetic potential $A(\theta, \phi, t)$.
A background rotation profile and toroidal magnetic field are imposed for all times and 
we study the behavior of the fields relative to these prescribed profiles.

The background differential rotation profile can be decomposed into a solid body rotation at frequency $\Omega$, and a latitudinal shear profile. Thus, in an absolute coordinate system, absolute vorticity $q$ is the sum of the generated vorticity, $\zeta^\prime$, the contribution from the shear profile, $\zeta_0$, and the Coriolis parameter, $f=2 \Omega \cos\theta$.
\begin{equation}
q = \zeta^\prime + \zeta_0 + 2 \Omega \cos\theta.
\end{equation}
The total magnetic potential, $A$, can be similarly decomposed into generated magnetic potential, $a$, 
and the background, $A_0$
\begin{equation}
A = a + A_0 .
\end{equation}
The relative vorticity is defined as
\begin{equation}
\zeta=\zeta^\prime + \zeta_0 = \nabla^2 \psi,
\end{equation}
which can also be represented in terms of the streamfunction $\psi$.

The evolution equations for $q$ and $A$ can then be written as
\begin{eqnarray}
\dfrac{\partial q}{\partial t} &=& J[q,~ \psi] + J[A,~ \nabla^{2}A] - \kappa \zeta^\prime - \nu_4 \nabla^{6} (\nabla^2 + 2) \zeta,
\nonumber \\
\dfrac{\partial A}{\partial t} &=& J[A,~ \psi] + \eta\nabla^{2} a ,
\label{EOM}
\end{eqnarray}
where $J[A,B]$ gives the Jacobian on the unit sphere:
\begin{equation}
J[A,~ B] \equiv \frac{1}{\sin\theta} \Big( 
\frac{\partial A}{\partial \phi} \frac{\partial B}{\partial\theta}
- \frac{\partial A}{\partial \theta} \frac{\partial B}{\partial \phi} \Big) .
\end{equation}
Note that, in contrast to previous work such as \citet{tobias2011}, the Lorentz force term in the momentum equation includes the prescribed field. This modification is necessary given the form of background field used.  Friction in the tachocline is parameterized with a frictional coefficient  $\kappa$.  To remove enstrophy as it cascades to small scales, hyperviscosity $\nu_4\nabla^6 (\nabla^2+2)  \zeta$ is included in the linear operator of Equation~(\ref{EOM}).  The appearance of the operator $(\nabla^2+2)$ ensures that the hyperviscosity does not change the total angular momentum. The use of hyperviscosity in place of ordinary viscosity has been used to model dynamics in the low magnetic Prandtl number limit appropriate to stellar interiors (and for liquid metals) \citep{seshasayanan}. Note however that we employ a regular diffusivity in the induction equation to maintain the balance between advection and diffusion there \citep{miesch_etal:2015}. 

We work on the unit sphere and in units of time such that $\Omega = 2 \pi$. A ``day'' is therefore a unit interval of time. The background vorticity field is chosen to have the form $\zeta_0(\theta, \phi) = \tilde{\zeta}_0~ Y_3^0(\cos \theta)$ and thus shears the flow symmetrically about the equator.   The imposed background toroidal potential is taken to be $A_0(\theta, \phi) = \tilde{A}_0~ Y_2^0(\cos \theta)$.   In all the simulations reported here (except for the linear stability analysis of Subsection \ref{stability}) we set $\tilde{\zeta}_0 = -3$ and $\tilde{A}_0 = 1/2$, in the joint instability regime where neither shear nor the background field by themselves suffice to trigger the instability.  We set the friction coefficient to be $\kappa = 0.05$, equivalent to an e-folding time of 20 days.  
 The hyperviscosity coefficient $\nu_4$ is chosen such that the most rapidly dissipating mode decays at a rate of $1$. 

\section{Methods}
\label{Methods}

In this section we give a brief description of the methods we utilize in the paper to analyze the behavior of the model, both in terms of the dynamics and statistics. These range from conservation laws,  linear stability analysis, fully nonlinear simulation (DNS) in both spectral and real space, partially nonlinear simulation (both quasilinear and generalized quasilinear) and statistical simulation. 

Our aim is to examine how well various approximations and statistical representations capture changes in the behavior of the model --- both in a dynamical and statistical sense.
  
 \subsection{Conservation Laws}
In the absence of damping and driving forces, the equations of motion (EOM) for the vorticity and magnetic potential conserve a number of linear and quadratic quantities.  In the pure hydrodynamic case ($A_0 = 0$), kinetic energy, enstrophy, and angular momentum are conserved.  Higher order Casimirs are also conserved in the continuum limit, but not in the finite-resolution numerical simulations that we employ. For $A_0 \neq 0$ the conserved quantities are 
angular momentum, total energy, mean square potential, and cross-helicity, the latter given by the average over the sphere of the product of the absolute vorticity and the scalar magnetic potential,
\begin{eqnarray}
\frac{1}{4 \pi} \int d^2\Omega~ q A \ .
\end{eqnarray}
These invariants are respected by the quasi-linear and generalized quasi-linear approximations \citep{marston2016}, as well as by the second-order cumulant expansion \citep{marston2014direct} that we shall utilize, and serve the practical purpose of validating code.  

\subsection{Linear Stability Analysis}
\label{stability}

The joint instability in a two-dimensional incompressible tachocline without viscous or ohmic dissipation was first described using a linear stability analysis \citep{GF97}. 
As described above, in our nonlinear equation solver, GCM, $\zeta_0$ and $A_{0}$ are set to be proportional to the spherical harmonics $Y_3^0$ and $Y_2^0$ respectively, for the sake of simplicity. Thus, $\zeta_0$ in GCM differs slightly from that used by Cally and others \citep{Cally_2001, GF97}. This is an acceptable substitution for our model, which is highly simplified and not physically precise. We have not been able to detect any qualitative difference in the simulations due to this modification. However, in order to confirm agreement between our work and earlier models, in this subsection we modify GCM to agree with the \citet{GF97} rotation profiles to check that we can reproduce the linear stability analysis.

The equations of motion described in Section \ref{Model} are rewritten in terms of equilibrium state stream functions in an absolute coordinate system. We first address the ideal case, in which friction, magnetic diffusivity, and hyperviscosity are set to zero.
We group together the Coriolis parameter, $f$, and the forcing function, $\zeta_0$, and represent them using a single streamfunction $\psi_0$.
\begin{equation}
q=\zeta^\prime+ (\zeta_0+ f)= \nabla^2 \psi^\prime + \nabla^2 \psi_0.
\end{equation} 
We define $\mu=\cos\theta$, and, assuming our prescribed fields are zonally symmetric, define a rotational angular frequency and Alfv\'en frequency for the prescribed fields. We set the frequencies to be the profiles given in \citet{GF97}. 
\begin{align}
\omega_0&=\frac{\partial \psi_0}{\partial \mu}= \Omega - \tilde{\zeta}_0 \mu^2 \\
\alpha_0&= \frac{\partial A_0}{\partial \mu}= \tilde{A}_0 \mu
\end{align}

The equations are linearized in the perturbations $\psi^\prime$ and $a$, with each field expanded in terms of associated Legendre Polynomials as 
$\psi^\prime= \sum\limits_{\ell=m}^\infty \Psi^{m}_\ell P^{m}_\ell(\mu) e^{i(m\phi- \omega t)}$ 
and 
$a= \sum\limits_{\ell=m}^\infty A^{m}_l P^{m}_\ell(\mu) e^{i(m\phi- \omega t)}$ where $\omega$ is the complex-valued angular frequency eigenvalue.  
The substitutions yield the equations:
 \begin{equation}
(m \omega_0 - \omega) \sum\limits_{\ell=m}^\infty A^{m}_\ell P^{m}_\ell=m \alpha_0 \sum\limits_{\ell=m}^\infty \Psi^{m}_\ell P^{m}_\ell(\mu) ,
\end{equation}
and
\begin{multline}
-(m \omega_0 - \omega) \sum\limits_{\ell=m}^\infty \ell(\ell+1)\Psi^{m}_\ell P^{m}_\ell-m \frac{d^{2}}{d\mu^{2}}[(1-\mu^{2})\omega_0]\sum\limits_{\ell=m}^\infty \Psi^{m}_\ell P^{m}_\ell+m \alpha_0\sum\limits_{\ell=m}^\infty \ell(\ell+1)A^{m}_\ell P^{m}_\ell \\+m \frac{d^{2}}{d\mu^{2}}[(1-\mu^{2})\alpha_0]\sum\limits_{\ell=m}^\infty A^{m}_\ell P^{m}_\ell=0 .
\end{multline}
Since the selected profiles for rotation and magnetic field are symmetric and antisymmetric about the equator respectively, either $\psi^\prime$ is symmetric and $a$ anti-symmetric, or $a$ is symmetric and $\psi^\prime$  antisymmetric. By truncating $\ell$ at a finite number $L_{max}$, two 2$L_{max}$ x 2$L_{max}$ matrices, one for the symmetric case and one for the antisymmetric case, can then be diagonalized numerically to find the growth rates of the instabilities.

\begin{figure}
\begin{center}
\includegraphics[width=14cm]{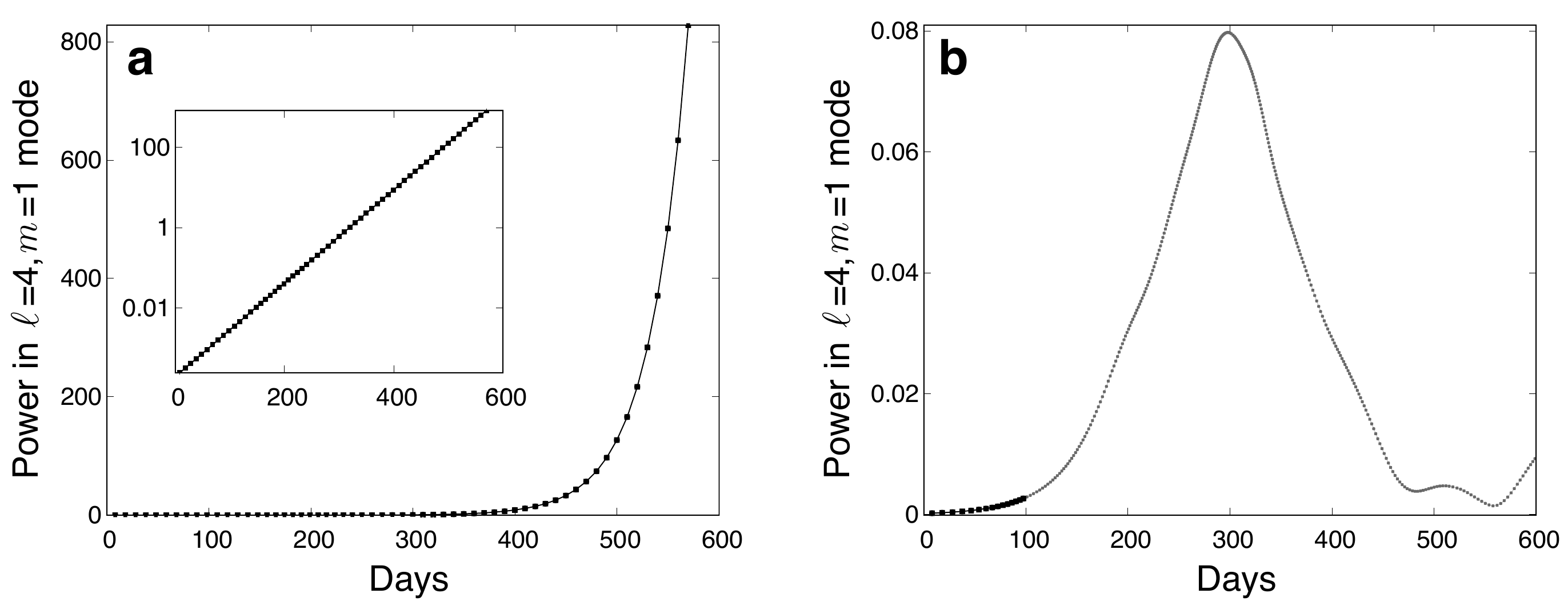}
\caption{\label{fig:linearnonlinear} Comparison between GCM and linear analysis for joint instability parameters with no dissipation. (a): Power in $\ell$=4, $m$=1 spherical harmonic over time in GCM code using prescribed field profiles from \citet{GF97} without nonlinear terms. An exponential fit with a growth rate of 0.027 is shown, as predicted by linear analysis. Inset shows same data on a semilog plot. (b): Power in $\ell$=4, $m$=1 mode over time in GCM, including nonlinear terms. Bolded black points show exponential fit with a growth rate of 0.027 for times before 100 days.}
\end{center}
\end{figure}

Using this method, the existence of the joint instability is confirmed in the dissipationless regime. The system continues to exhibit an instability if dissipation is added to our analysis (nonzero friction and/or magnetic diffusivity), provided the dissipation is not too strong. As expected, increasing dissipation decreases the growth rate of the instability \citep{dikpati2004linear}. For all values of parameters examined, $m=1$ is the most unstable mode, as found previously \citep{GF97, Cally_2001}. Additionally, GCM with appropriately modified profiles is found via timestepping to reproduce the linear analysis both when nonlinear terms are turned off, and in the limit of small amplitude waves with nonlinear terms included (Figure \ref{fig:linearnonlinear}).

\subsection{Direct Numerical Simulation in Real and Spectral Space}

Fully nonlinear (NL) spectral DNS with truncation $0 \leq \ell \leq L$ and $|m| \leq \min\{\ell, M\}$ is performed.  We set spectral cutoffs $L = 60$ and $M =15$ and verify, by comparison with high-resolution simulations carried out in real space, that these cutoffs suffice.  
The pure spectral version of Equation~(\ref{EOM}) is integrated forward in time using a fourth-order-accurate Runge-Kutta algorithm with an adaptive time step $\Delta t$.  Each time step requires ${\cal O}(L^3 M^2)$ floating point computations that at high resolutions would be prohibitively expensive compared to a pseudo-spectral algorithm but is feasible here for the moderate resolutions that we study.  The computation is made faster by skipping over triads that vanish due to symmetry.  

To verify that the full spectral simulation has sufficient resolution, finer-scale DNS of the fluid is also performed in real space on a spherical geodesic grid \citep{Dritschel:2015ei} of D = 163,842 cells; the lattice operators conserve energy and enstrophy. The vorticity evolves forward in time by a second-order accurate leapfrog algorithm, with a Robert-Asselin-Williams filter of $0.001$ and $\alpha = 0.53$ \citep{Williams:2009ix}.  The time step is fixed at $\Delta t = 0.003$.  

\subsection{QL and GQL Numerical Simulation}

\begin{figure}
\begin{center}
\includegraphics[width=5cm]{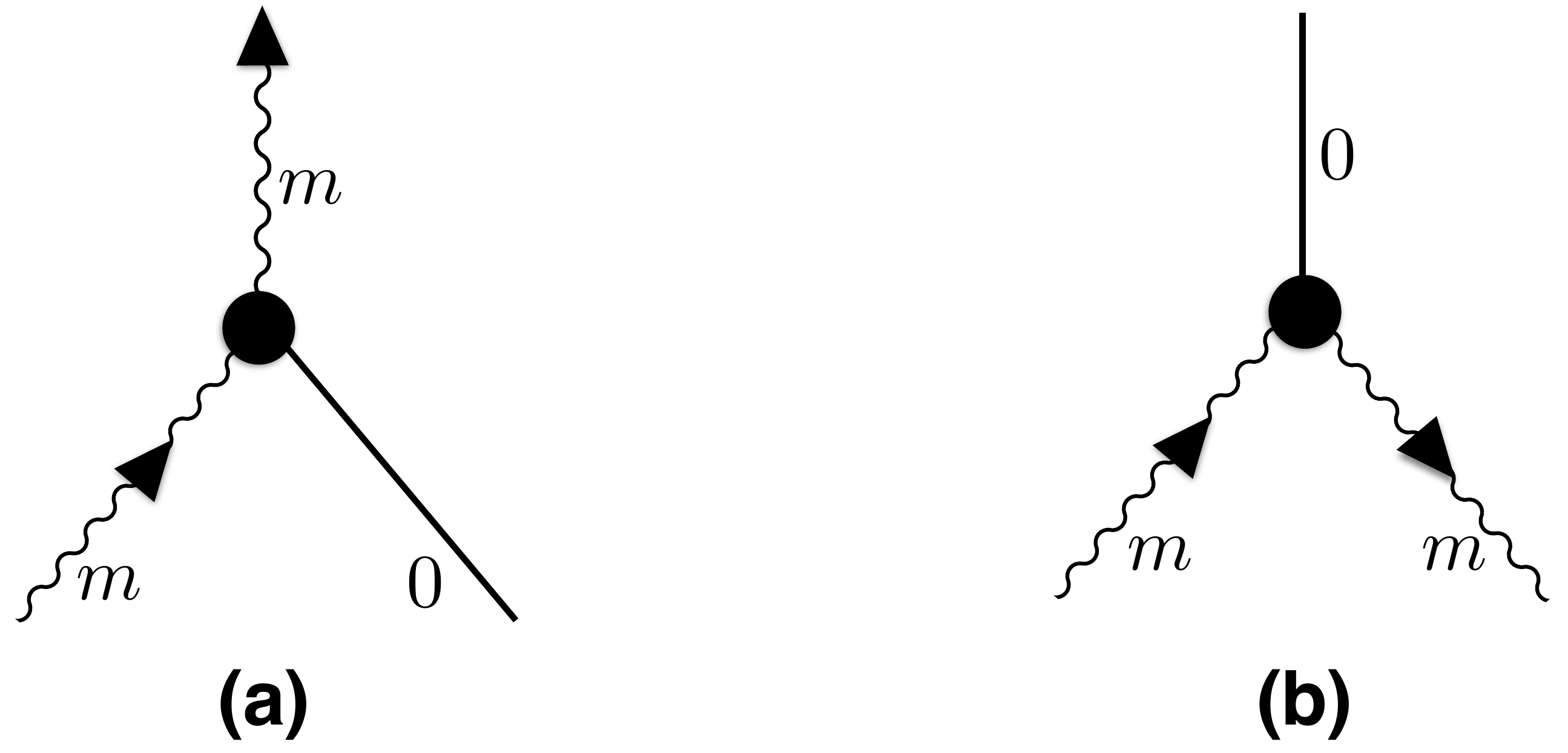}
\caption{\label{fig:QLtriads} Triadic interactions retained in the QL approximation.  The zonal wavenumber of the eddies is labeled by $m$; the zonal mean flow has $m = 0$ and is depicted by the solid line. (a) An eddy scatters off the zonal mean flow. (b) Two eddies of equal but opposite zonal wavenumber combine to modify the zonal mean flow.}
\end{center}
\end{figure}

The quasi-linear (QL) approximation has its historical origins in the work of \cite{MALKUS:1954dh} and \cite{Herring1963}, with perhaps the clearest earliest exposition given in the plasma context by \citet{VVS63}.  The approximation is a self-consistent mean-field theory that in this context retains the scattering of eddies (defined as perturbations with respect to the zonal mean) off the mean zonal flow, and processes in which two eddies of equal and opposite zonal wavenumber combine to influence the mean flow.  The two retained triadic interactions are depicted in Fig. \ref{fig:QLtriads}.  Note that the general eddy + eddy $\rightarrow$ eddy scattering is dropped.  

\begin{figure}
\begin{center}
\includegraphics[width=14cm]{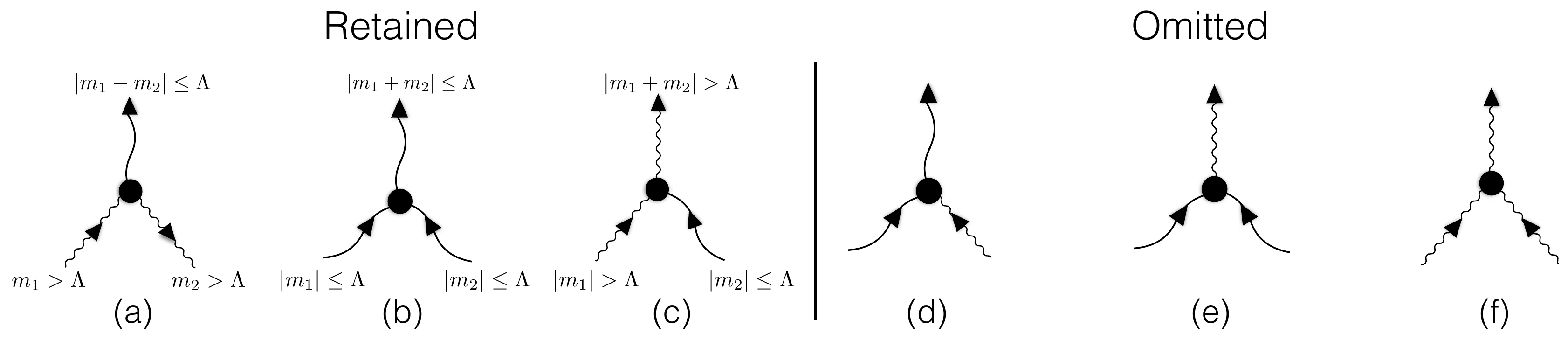}
\caption{\label{fig:GQLtriads} Triadic interactions retained and omitted in the GQL approximation.  Modes with low zonal wavenumbers $|m| \leq \Lambda$ interact fully nonlinearly (b).  The interactions of the low modes with the high wavenumber modes, (a) and (c), generalizes the quasi-linear interactions shown in Fig. \ref{fig:QLtriads}.}
\end{center}
\end{figure}

A generalization of QL called the generalized quasi-linear approximation (GQL) \citep{marston2016}  systematically improves upon the QL approximation by including, self-consistently, fully nonlinear interactions among large-scale modes.  The zonal mean of the QL approximation is generalized to  include modes with low zonal wavenumber at or below a cutoff $|m| \leq \Lambda$; see Fig. \ref{fig:GQLtriads}.  $\Lambda = 0$ is QL, while the limit $\Lambda = M$ recovers the full nonlinear EOMs; thus GQL interpolates between QL and NL.  \citet{marston2016} examined GQL in the context of two-dimensional barotropic turbulence on a spherical surface and on a $\beta$-plane and showed it to be more effective than the QL approximation in reproducing both the dynamics and statistics of these flows away from equilibrium. Remarkably, this remained true even if only a single extra mode was retained in the large scales.  Subsequent work demonstrated the effectiveness of the GQL approximation for the case of axisymmetric (2D) helical magnetorotational instability (HMRI) at low magnetic Reynolds number \citep{Child:2016go}, and in 3D rotating plane Couette flow \citep{Tobias:2016kp}.  

\subsection{Direct Statistical Simulations: CE2}
\subsubsection{CE2 simulations}
In order to understand better the physical mechanism behind the dynamics and statistical properties of the model, a Direct Statistical Simulation (DSS) technique based upon expansion of the equations of motion for the equal-time cumulants is also used.  We retain only the first and second cumulants (CE2); for a review, see \citet{marston2014direct}.  In the CE2 approximation, the contribution of the third cumulant to the tendency of the second cumulant is neglected, an approximation that is mathematically equivalent to the QL approximation of eliminating the triadic interaction between two eddies that results in a third eddy. The CE2 equations are an exact closure within the QL approximation.
DSS is best suited to describing systems with large-scale inhomogeneous and anisotropic flows.  The approach was applied to a different, stochastically-forced, tachocline model in \citet{tobias2011}.  Here we apply it to the deterministic Cally model.  

A program that implements spectral DNS, real-space DNS, QL, GQL, and DSS, and also includes all the graphical tools needed to visualize statistics, is freely available\footnote{The application ``GCM'' is available for macOS on the Apple Mac App Store at URL http://appstore.com/mac/gcm}.  The Objective-C++ and Swift programming languages are employed. C blocks and Grand Central Dispatch enable the efficient use of multiple CPU cores.  

\section{Results}
\label{Results}

\begin{figure}
\begin{center}
\includegraphics[width=14cm]{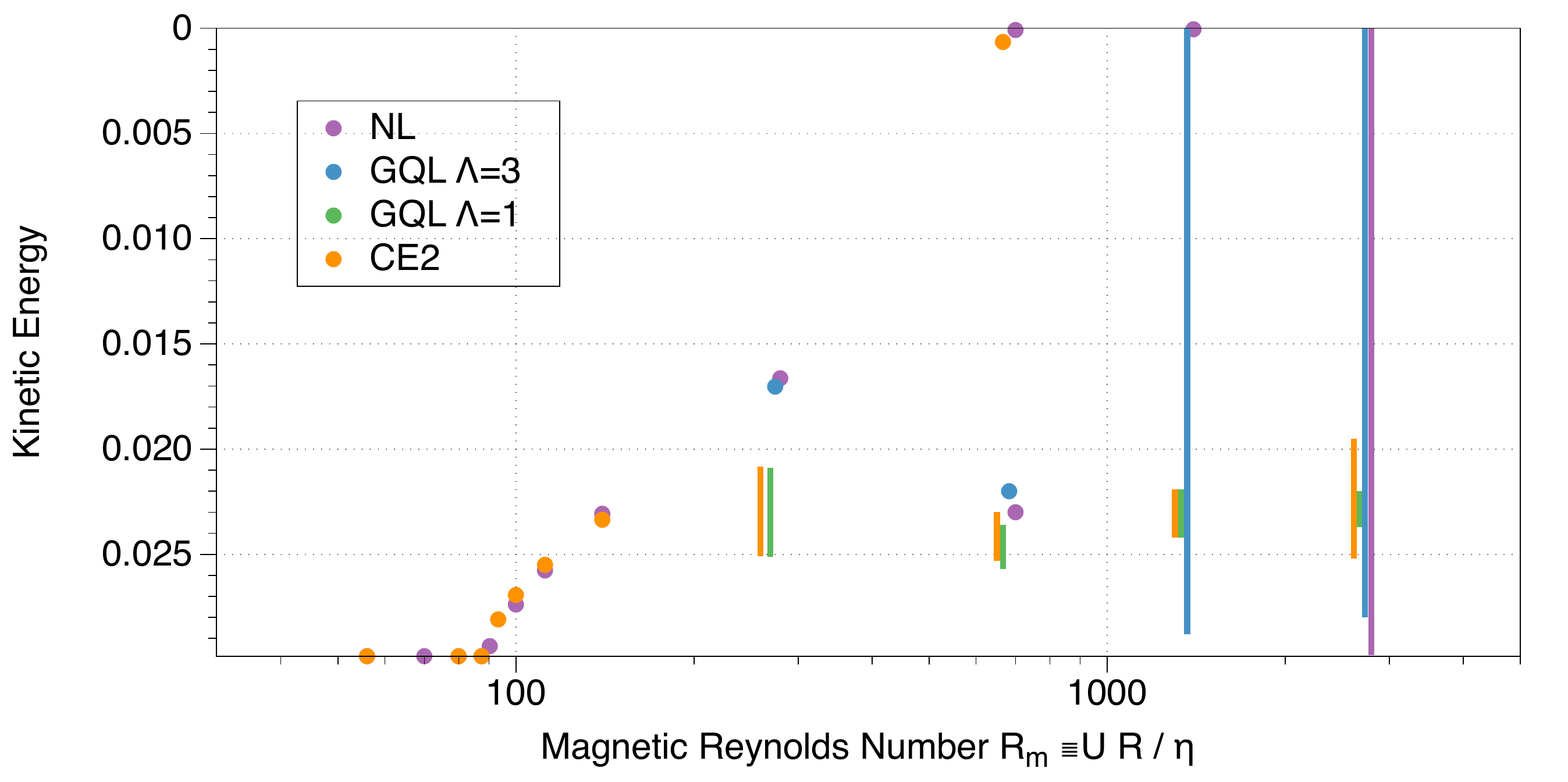}
\caption{{\label{fig:bifn} Bifurcation Diagram showing the nature of the solutions obtained by the various methods as a function of magnetic Reynolds number $R_m$. Note the kinetic energy axis has been inverted so that solutions with the kinetic energy of the basic state appear at the bottom of the diagram.  The data has been shifted slightly in $R_m$ for clarity.
  }}
\end{center}
\end{figure}

\begin{figure}
\begin{center}
\includegraphics[width=8.5cm]{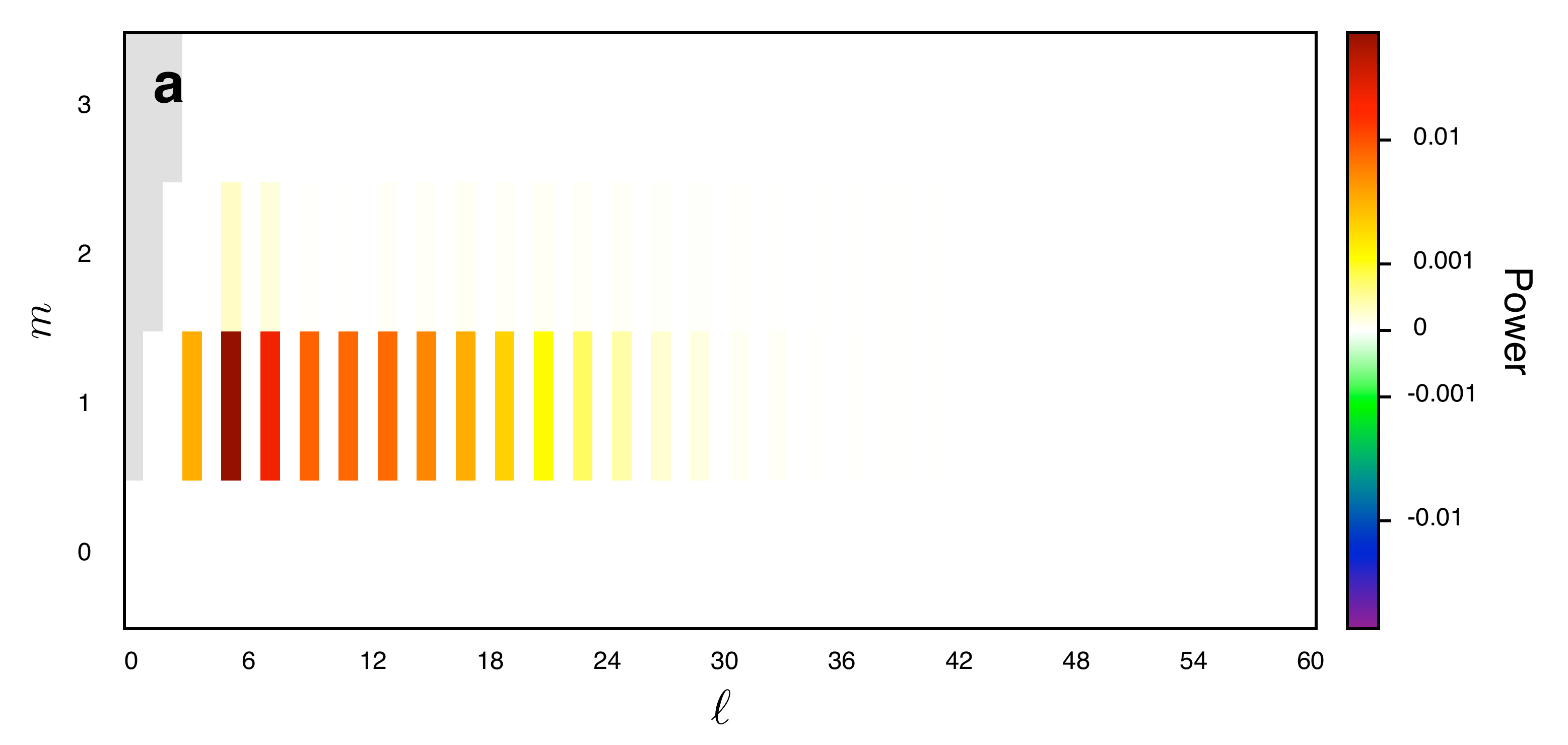}
\includegraphics[width=8.5cm]{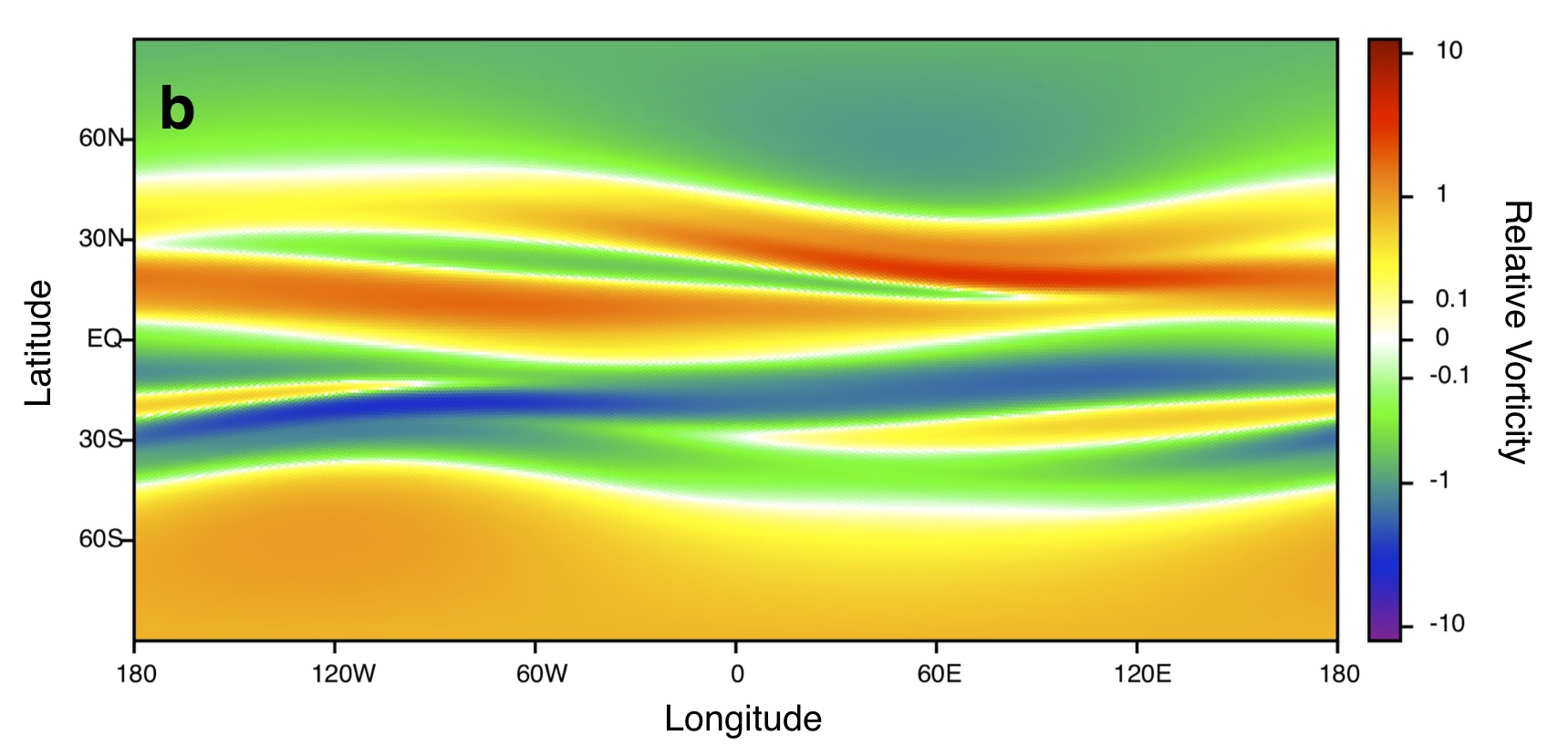}
\includegraphics[width=8.5cm]{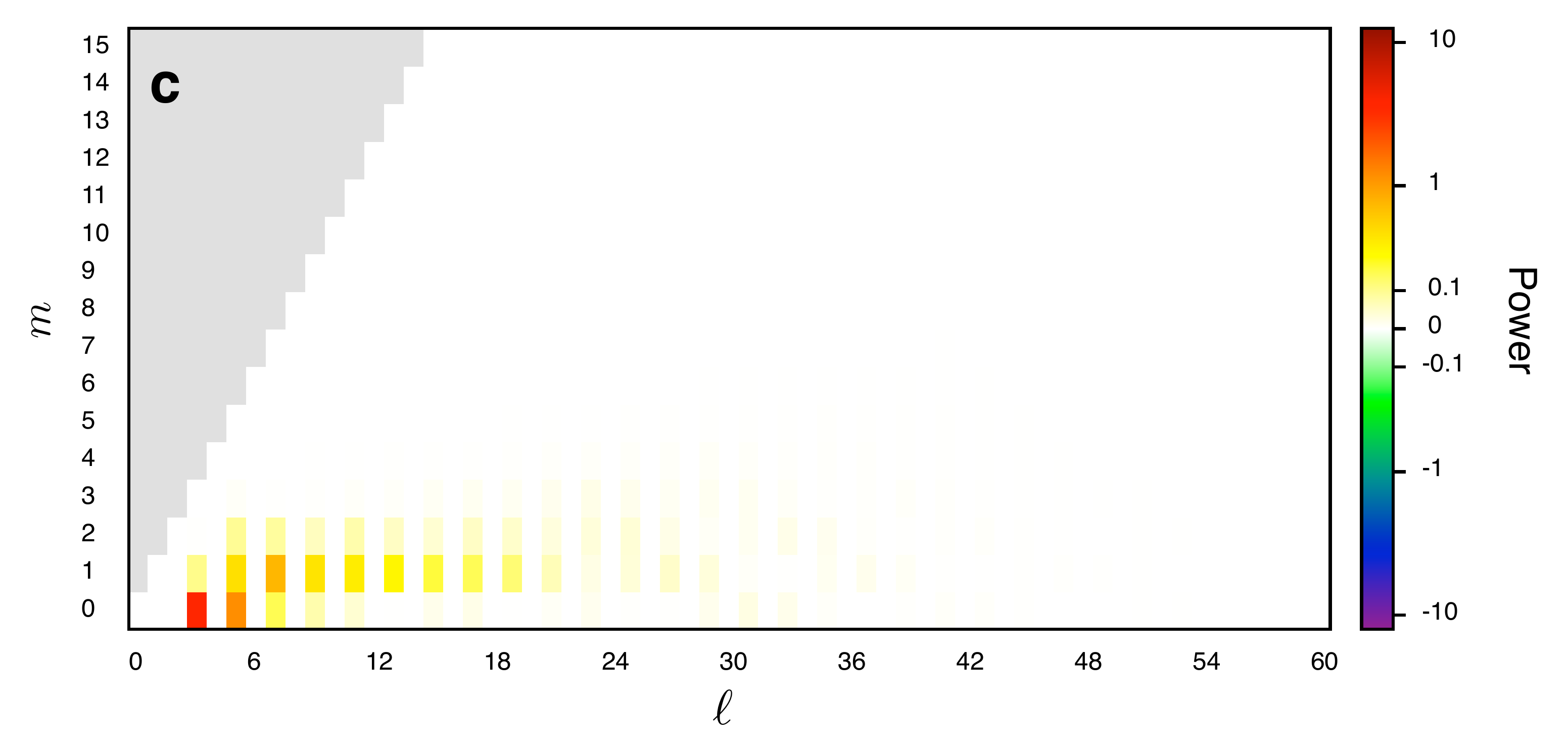}
\caption{{\label{fig:power_spectra} (a) Power spectrum for $R_m=90$ -- energy is largely restricted to the $m=1$ mode. Here $l$ is the spherical wavenumber and $m$ is the zonal wavenumber.  (b) Snapshot of the relative vorticity and (c) power spectrum for $R_m\sim 300$. For this value of $R_m$, more azimuthal structure is observed.
  }}
\end{center}
\end{figure}

\begin{figure}
\begin{center}
\includegraphics[width=8.5cm]{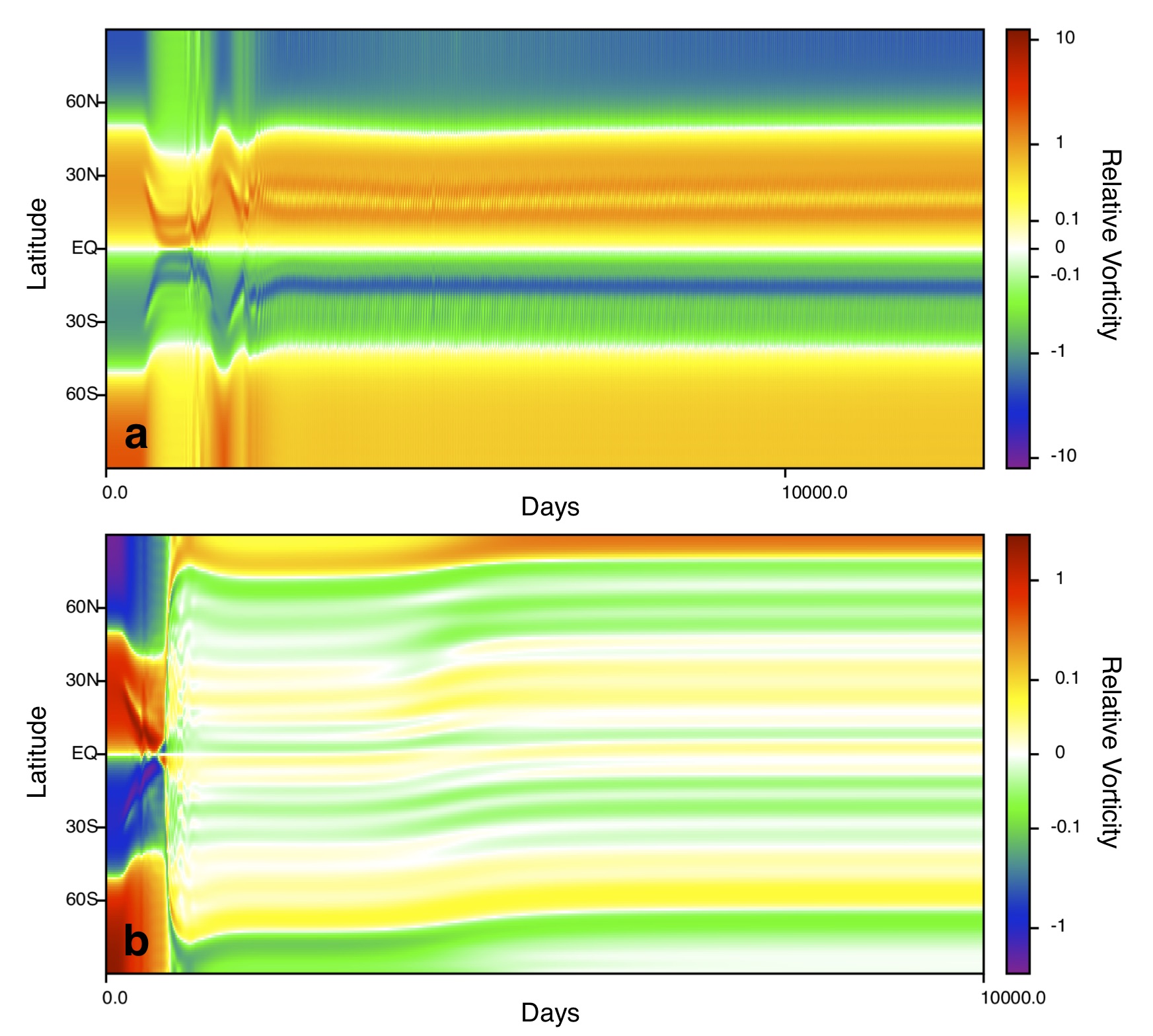}
\caption{{\label{fig:Relvort_hys} Hysteresis in the fully nonlinear simulations at an intermediate value of the magnetic Reynolds number. Timelines of the zonally averaged relative vorticity for $R_m\sim 700$ show (a) the high vorticity solution and (b) the low vorticity solution.
  }}
\end{center}
\end{figure}

We consider the evolution of the instability and determine how the dynamics change as the magnetic Reynolds number is increased, for fixed profiles of differential rotation and magnetic field and fixed relaxation time. The magnetic Reynolds number is defined as
\begin{equation}
R_m = \dfrac{U R}{\eta},
\end{equation}
where $U$ is the zonal velocity of the basic state at the equator, $R$ is the radius, and $\eta$ is the magnetic diffusivity. The results are summarized in the bifurcation diagram of Figure~\ref{fig:bifn}, where we use the kinetic energy as a fiducial measure of the amplitude of the solution. We note that this diagram has been flipped such that solutions with the kinetic energy of the basic differential rotation appear at the bottom, so that the figure resembles a traditional bifurcation diagram. We begin by describing the transitions in behavior as $R_m$ is increased as determined by the fully nonlinear simulations. We will then go on to determine how well these transitions are captured by the various approximations (CE2, GQL $\Lambda=1$, and GQL $\Lambda=3$) subsequently.

\subsection{Fully Nonlinear Dynamics}

For low $R_m$ ($ < 90$), the basic state is stable, as described by the linear theory above.  For $R_m>90$, NL solutions (denoted in purple) are initially located on a primary branch, where the energy in the perturbation is weak and contained almost completely in the $m=1$ mode, as shown by the power spectrum in Figure~\ref{fig:power_spectra}(a). As $R_m$ is increased, the solution remains on this primary branch, with the solution having a characteristic $m=1$ dependence on longitude and a banded dependence on latitude (as shown in  Figure~\ref{fig:power_spectra}(b); this solution does have power in higher azimuthal wavenumbers, though these are limited to $m \lesssim 3$ as shown in Figure~\ref{fig:power_spectra}(c).

At $R_m\sim700$, our NL simulation is able to find two different types of solution depending on initial conditions and bistability is present. Figure~\ref{fig:Relvort_hys} shows the timelines of the zonally averaged relative vorticity for two different initial conditions. It can be seen clearly that there are two different states, one of which, shown in Figure~\ref{fig:Relvort_hys}(a) has a high relative vorticity (as for those at lower $R_m$) and one (shown in Figure~\ref{fig:Relvort_hys}(b)) yields a state with much lower vorticity.

\begin{figure}
\begin{center}
\includegraphics[width=14cm]{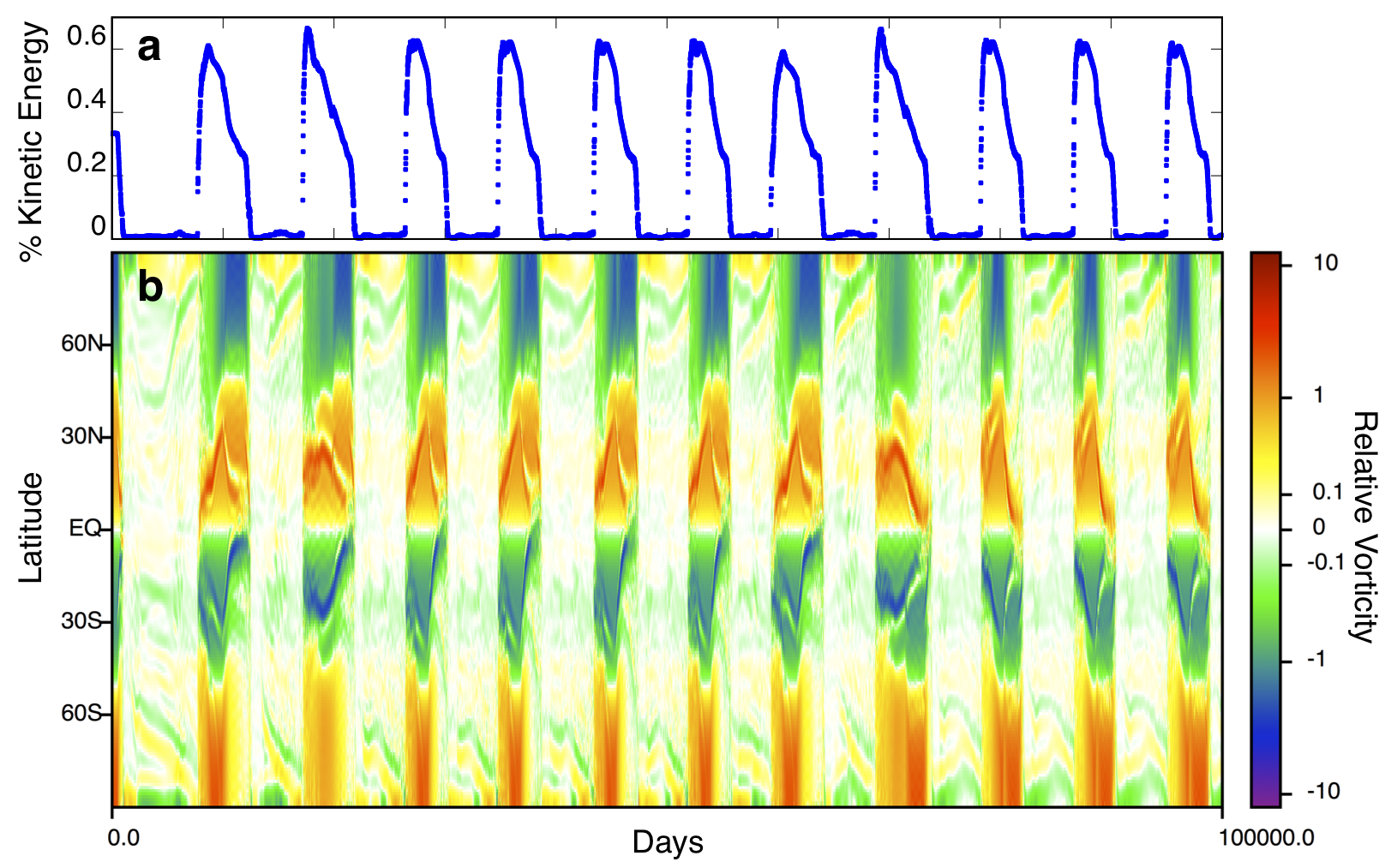}
\caption{{\label{fig:vorttimeline} Abrupt transitions in relative vorticity ($\zeta$) for a spectral simulation with $L_{max}=60$, $M_{max}=15$, and $R_m=2800$. (a): Fraction of total energy in the form of kinetic energy versus time. (b): Zonally averaged relative vorticity for the same time period. The kinetic energy share tracks the transitions between the two distinct states and highlights the rapidity of the switch.%
}}
\end{center}
\end{figure}

\begin{figure}
\begin{center}
\includegraphics[width=14cm]{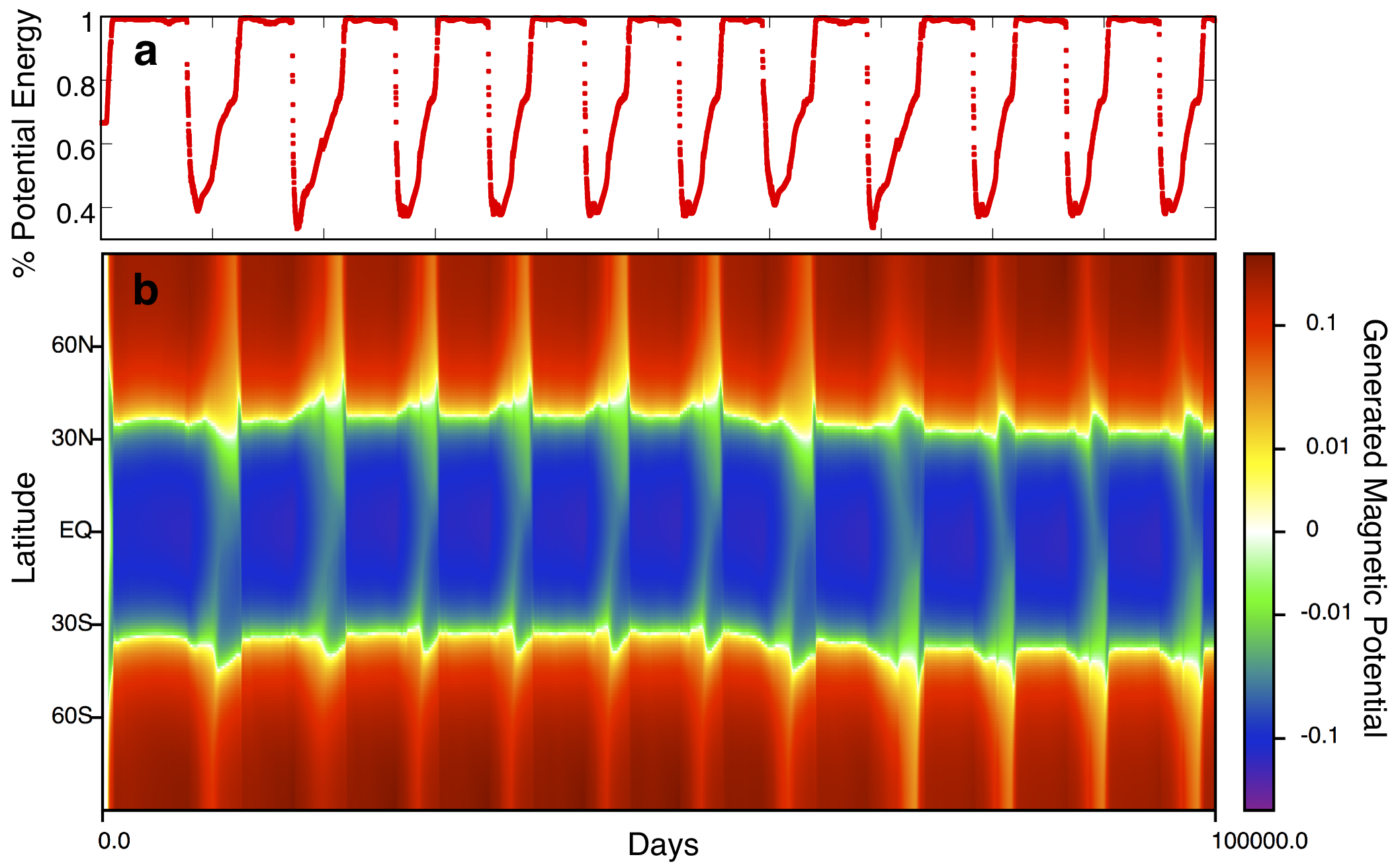}
\caption{{\label{fig:pottimeline} Abrupt transitions in generated magnetic potential, $a$, for a spectral simulation with $L_{max}=60$, $M_{max}=15$, and  $R_m=2800$. (a): Fraction of total energy in the form of magnetic potential energy versus time. (b): Zonally averaged generated magnetic potential for the same time period.
}}
\end{center}
\end{figure}

\begin{figure}
\begin{center}
\includegraphics[width=14cm]{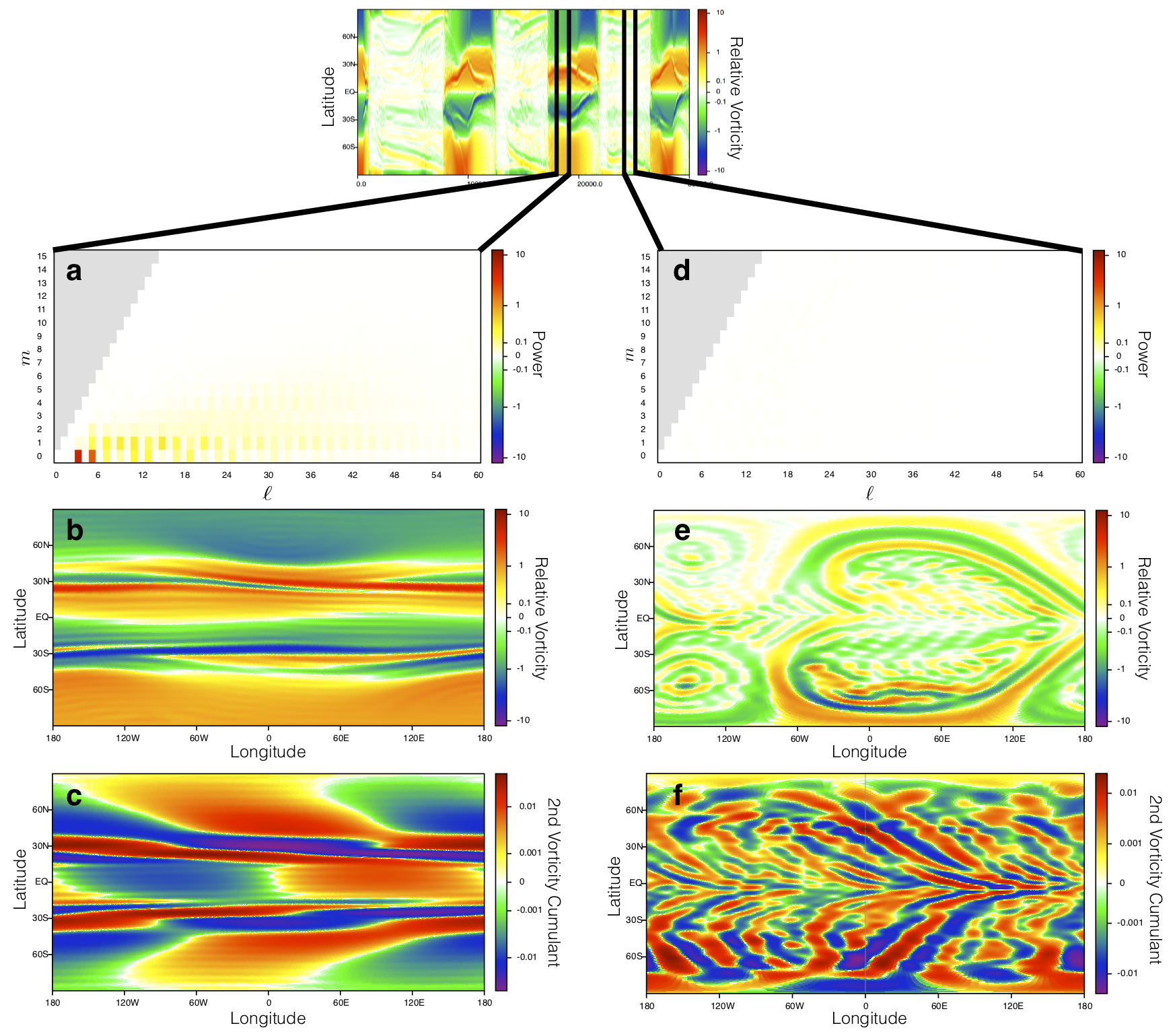}
\caption{{\label{fig:spectralsnaps} Snapshots of both high and low vorticity states for $R_m=2800$. In the top timeline, the double bars at 18,000-19,000 and 24,000-25,000 days show the period over which time averaging is performed for Figure \ref{fig:compare}. The two columns of figures are representative snapshots taken within these two windows respectively (not time averaged). (a) and (d): Power spectra of vorticity modes. (b) and (e): Relative vorticity on a cylindrical projection. (c) and (f): Second vorticity cumulant on a cylindrical projection.
}}
\end{center}
\end{figure}

For the highest value of $R_m$ considered ($R_m=2800$), interesting relaxation dynamics are found as shown in Figures \ref{fig:vorttimeline}, \ref{fig:pottimeline}, and \ref{fig:spectralsnaps}. From these figures it is clear that the relaxation occurs as an oscillation between the high and low vorticity states described in the hysteretic scenario above. Both of these states are now unstable at this high $R_m$ (presumably having undergone Hopf bifurcations) and the large amplitude relaxation oscillation takes the form of a near-heteroclinic solution oscillating between the two unstable states. The transitions between these two states occur rapidly, on a timescale of approximately 300 days, with a cycle period of approximately 9,000 days. Relative vorticity in the high vorticity state is approximately zonally symmetric and power is distributed among a broad range of spherical harmonics. In the low vorticity state, the relative vorticity snapshot shows fine structure and very little power in relative vorticity spherical harmonics (see Figure \ref{fig:spectralsnaps}). 

\begin{figure}
\begin{center}
\includegraphics[width=14cm]{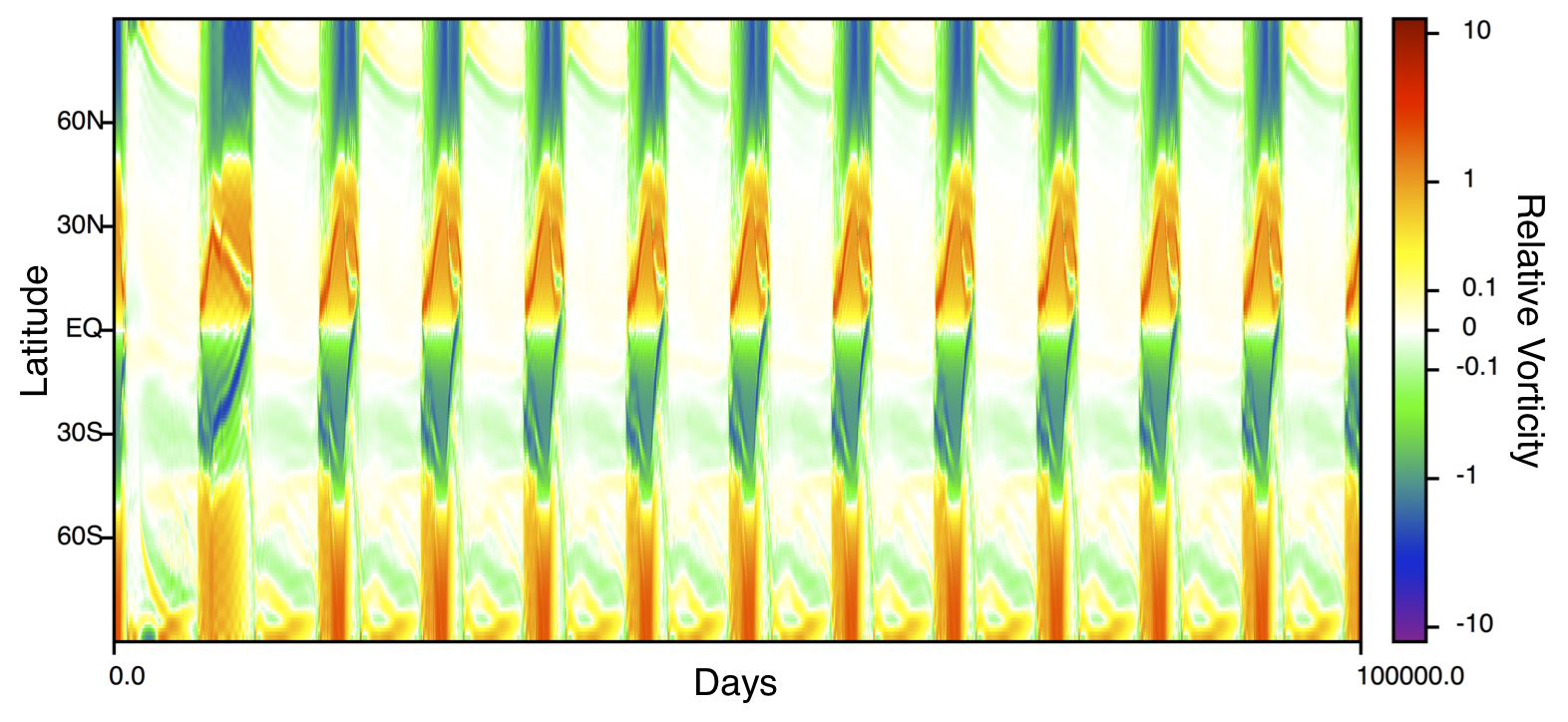}
\caption{{\label{fig:geovsspectral} A real space simulation using the same parameters as the spectral space simulation of relative vorticity in Figure 7 ($R_m=2800$). The behavior is qualitatively similar. 
}}
\end{center}
\end{figure}

We note that this behavior is robust. It appears in both spectral and geodesic simulations. The transitions do not change qualitatively as resolution in real space is increased, and, as seen in Figure \ref{fig:geovsspectral}, the transitions are reproducible using both a fully spectral simulation in a basis of spherical harmonics with $L_{max}=60$ and $M_{max}=15$, as well as a real space simulation using six (pictured) and seven geodesics. The transitions are consistent and regular even in our longest, highest resolution runs.  The addition of small stochastic forcing of the vorticity also does not alter the behavior, demonstrating that the transitions are not triggered by noise.  

\begin{figure}
\begin{center}
\includegraphics[width=14cm]{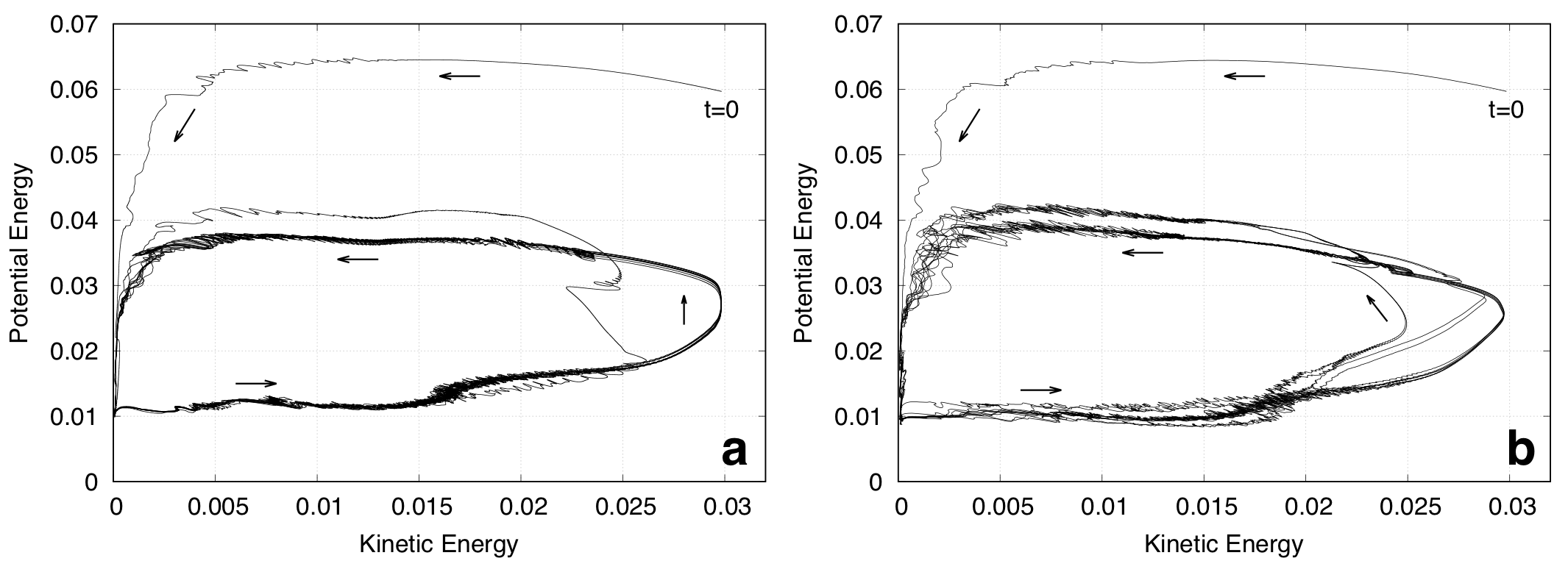}
\caption{{\label{fig:cyclegeovsspectral} Abrupt transitions in kinetic energy-potential energy space for $R_m = 2800$. (a): Real space simulations. (b): Spectral space simulations. Arrows show the direction of time. Both simulations reach a steady cycle after a brief initialization period. 
}}
\end{center}
\end{figure}

Perhaps it is most informative to describe this relaxation oscillation in the `potential energy-kinetic energy' phase plane. In Figure \ref{fig:cyclegeovsspectral}, the abrupt transitions appear as a near-elliptical path. The low vorticity state occurs along the y-axis, and the high vorticity state occurs along the rest of the path. In one period of the relaxation oscillation, the system moves slowly down the y-axis and then much more quickly along the rest of the ellipse, such that the time spent in both states is comparable, as we see in Figure \ref{fig:vorttimeline}.  

\begin{figure}
\begin{center}
\includegraphics[width=14cm]{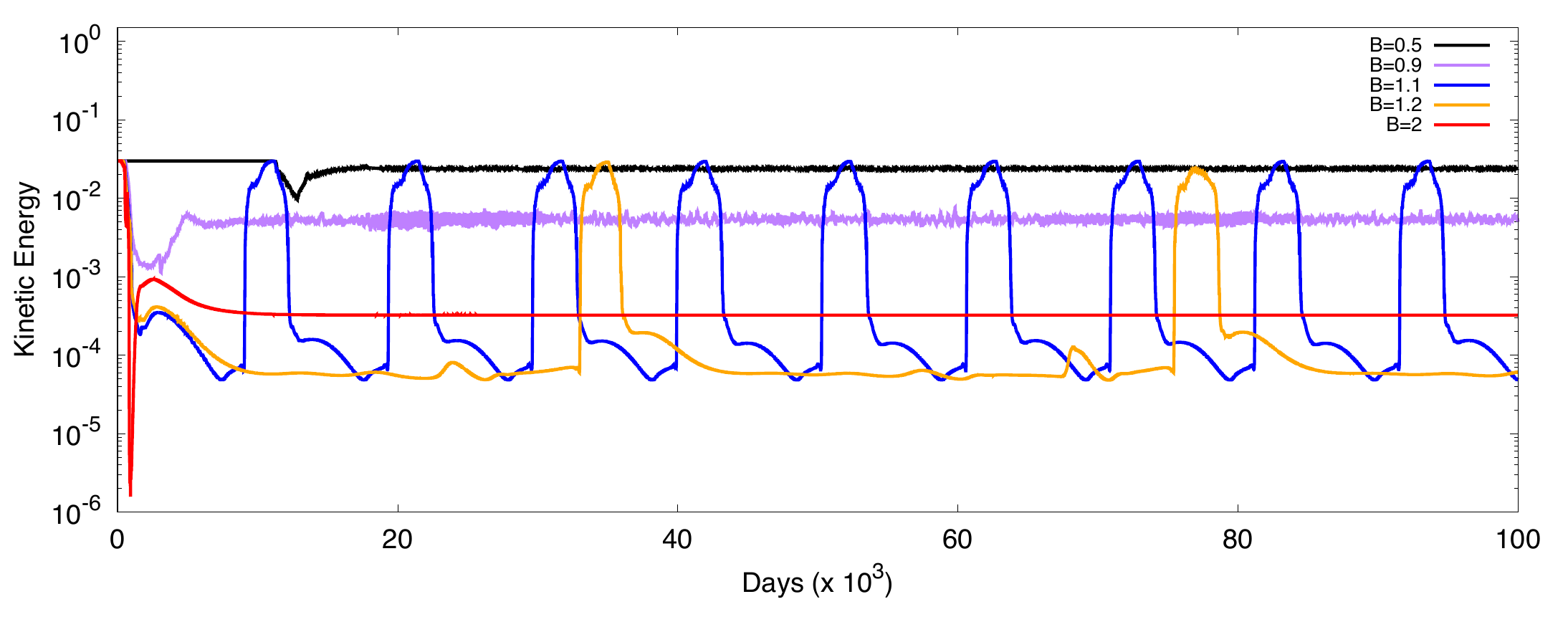}
\includegraphics[width=14cm]{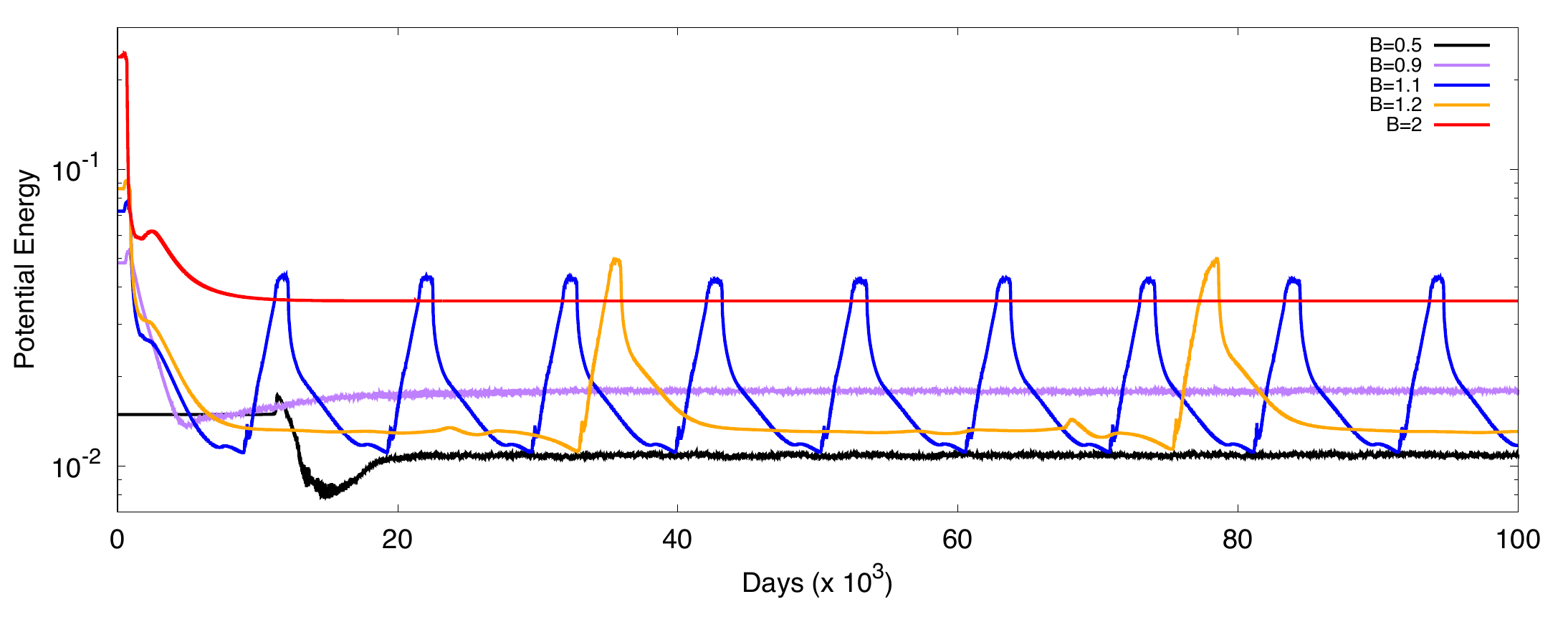}
\caption{{\label{fig:dependence_on_B} Abrupt transitions in kinetic energy-potential energy space for $R_m=2800$ and varying applied field $B$. (a): Kinetic Energy versus time (b): Potential Energy versus time
}}
\end{center}
\end{figure}

The physics of this relaxation oscillation can be elucidated by performing additional numerical simulations for varying applied magnetic field strengths $B$ (at fixed $R_m$). Figure~\ref{fig:dependence_on_B} shows both the birth and death of the relaxation oscillation as $B$ is varied. For small $B=0.5$ or $0.9$, the solution takes the form of a high vorticity state (with large kinetic energy and small kinetic energy), whilst for high $B=2$ it takes the form of a low vorticity state. For intermediate values of $B$ the solutions has the character of a relaxation oscillation. These additional simulations also show that the period of the oscillation is controlled by the near heteroclinic nature of the solution --- as $B$ is increased the period increases as the solution spends increasing amount of time near the invariant low vorticity state. This behaviour is reminiscent of other ``Limit Cycle Oscilations'' (or LCOs) in plasma such as Edge Harmonic Oscillation (EHO) that is present owing to the action of an instability (the kink-peeling instability) in the presence of a sheared rotation (in this case the $\boldsymbol{E} \times \boldsymbol{B}$ rotation), see, for example, \citet{Wilksetal2018}.

\subsection{CE2 and GQL}

We now turn to how well the DSS method at CE2 and the DNS methods utilizing the GQL approximations with various severe truncations ($\Lambda=1,3$) perform. Figure~\ref{fig:bifn} gives the average kinetic energy of the solutions found by these methods as a function of $R_m$. We note that the DSS and GQL runs were performed at the same $R_m$ as the corresponding NL solution though the results have been shifted slightly in the bifurcation diagram for ease of viewing.

\begin{figure}
\begin{center}
\includegraphics[width=14cm]{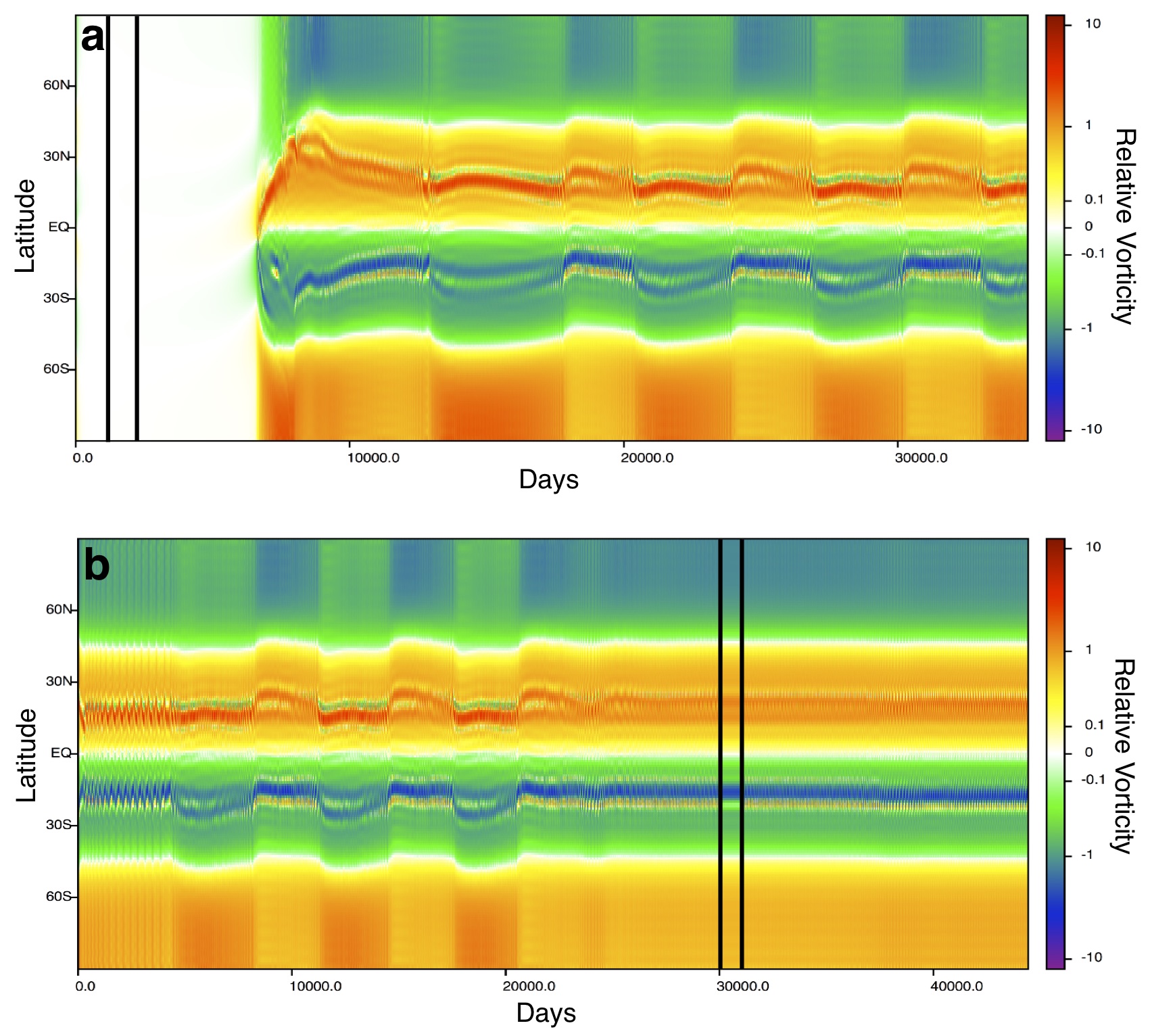}
\caption{{\label{fig:CE2} Zonally averaged relative vorticity versus time for CE2 simulations at $R_m=2800$. (a): CE2 initialized with perturbations in magnetic potential. (b): CE2 initialized with perturbations in vorticity. Time averaging is performed for the 1,000 day period between the double bars in both figures for Figure \ref{fig:compare}.
}}
\end{center}
\end{figure}

We begin by examining the efficacy of the most severe quasilinear approximation CE2. Figure~\ref{fig:bifn}  shows that CE2 captures the linear instability (as it must) and the amplitude of the primary branch, with the orange circles tracking the purple circles for $R_m<200$. This is encouraging for quasilinear theory. However by $R_m\sim300$  discrepancies occur between CE2 and NL. Here CE2 finds an oscillatory low vorticity state (whereas NL found a steady high vorticity state as described above). Interestingly, subsequent  increase of $R_m$ to $\sim 700$ increases the accuracy of CE2. It is remarkable that it is able to reproduce the hysteresis found by NL (though it finds an oscillatory rather than a steady low vorticity state). However, perhaps unsurprisingly, for the highest $R_m$ considered CE2 is unable to reproduce the relaxation oscillations shown by NL (Figure \ref{fig:CE2}). When initialized with sufficient magnetic potential energy, CE2 passes close to a state with very low vorticity and then makes a rapid transition to an periodic solution that is statistically similar to the high vorticity mode seen in NL trials (Figure \ref{fig:CE2}(a)). After this first transition, the system remains in this state. When only the vorticity field is perturbed during initialization, CE2 immediately reaches the approximate fixed point (Figure \ref{fig:CE2}(b)). Interestingly, this transition breaks north-south symmetry imposed at initialization when it reaches the high vorticity mode. 

\begin{figure}
\begin{center}
\includegraphics[width=14cm]{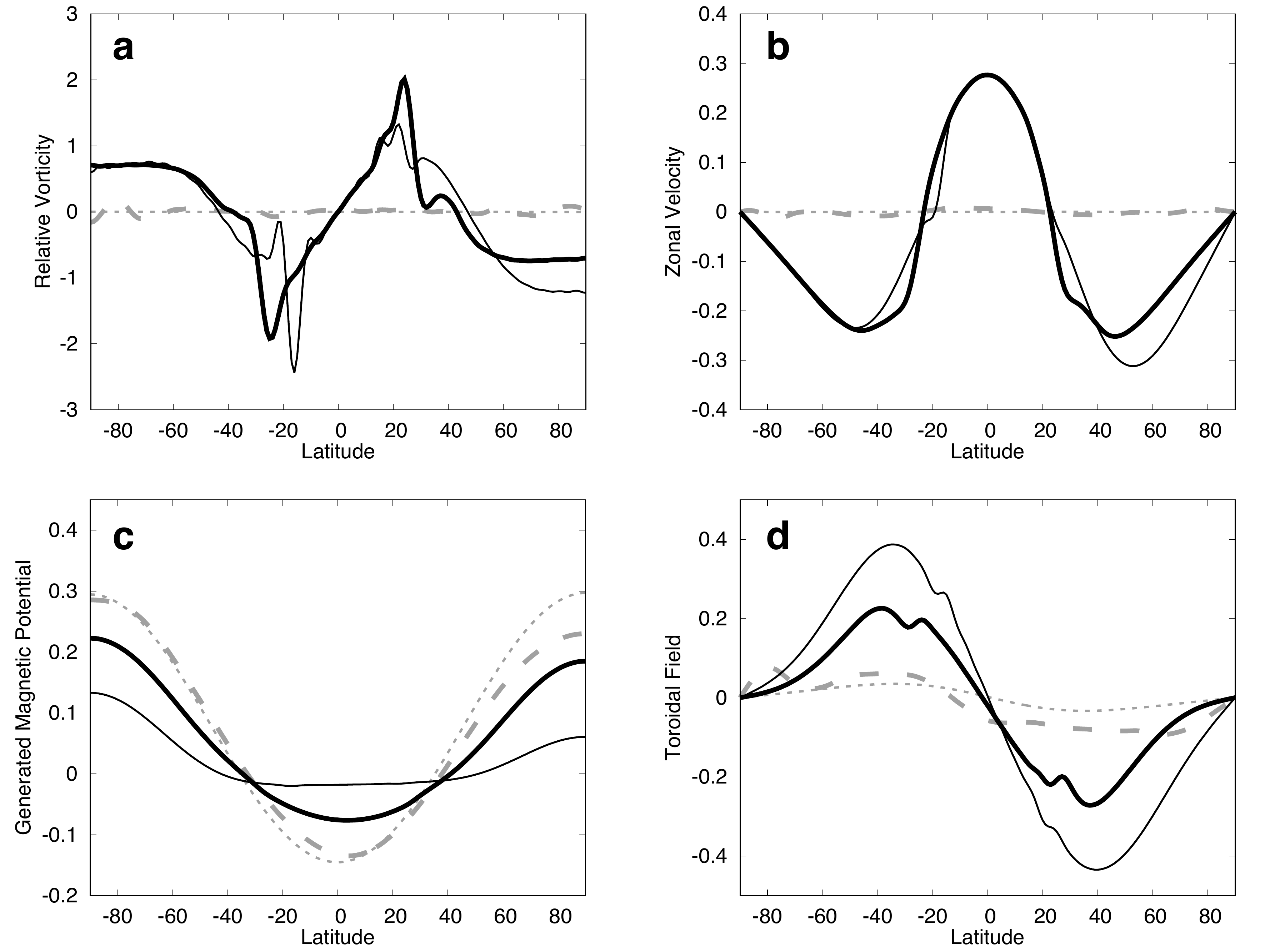}
\caption{{\label{fig:compare} Comparison between time averaged field profiles for $R_m=2800$. Thick lines correspond to fully nonlinear spectral DNS, and thin lines correspond to CE2. The thick solid line is the first time averaged period (high vorticity) in Figure \ref{fig:vorttimeline}, and the thick dashed line is the second period (low vorticity) in Figure \ref{fig:vorttimeline}. The thin dashed line is the time averaged portion of Figure \ref{fig:CE2}a, and the thin solid line is the time averaged portion of Figure \ref{fig:CE2}b. Each plot compares these four time averaged and zonally averaged fields, with (a) comparing relative vorticity, (b) comparing zonal velocity (not including solid body rotation), (c) comparing generated magnetic fields, and (d) comparing toroidal fields. Fields are plotted as a function of latitude.%
}}
\end{center}
\end{figure}

Time averaging is performed in both the low vorticity transient state and for the final oscillatory state. The zonally averaged fields are compared with time averaged results from spectral NL for high and low vorticity states, as shown in Figure \ref{fig:compare}. The NL low vorticity state and the CE2 low vorticity state have statistically similar field profiles, while the NL high vorticity state and the CE2 high vorticity fixed point also have statistically similar field profiles. Thus, CE2 is capable of accessing both of the states that the NL simulation transits between, but is incapable of reproducing the periodic transitions between them.

\begin{figure}
\begin{center}
\includegraphics[width=14cm]{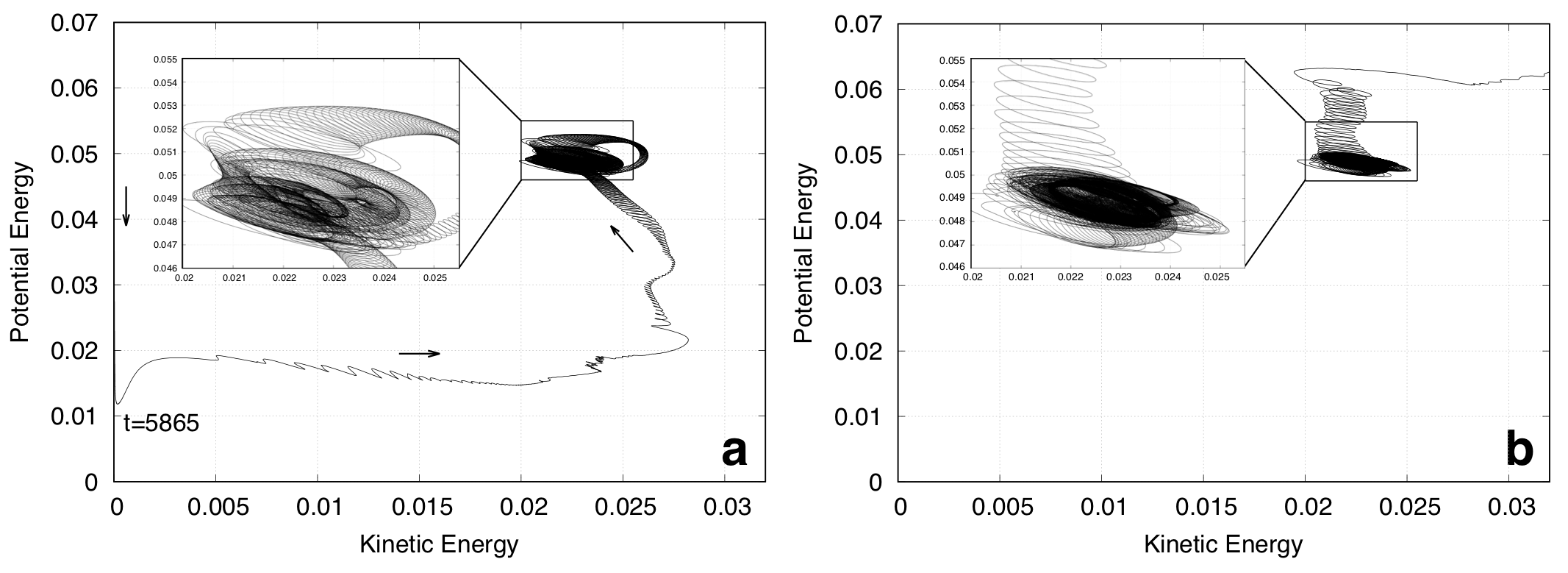}
\caption{{\label{fig:cycleCE2} CE2 simulations in kinetic energy-potential energy space for $R_m=2800$. Arrows show the direction of time, and the insets show the long time behavior. (a): CE2 initialized with perturbations in magnetic potential, as in Figure \ref{fig:CE2}a.  (b): CE2 initialized with perturbations in vorticity, as in Figure \ref{fig:CE2}b. 
}}
\end{center}
\end{figure}

This interpretation agrees well with the time series viewed in energy space. In contrast to Figure \ref{fig:cyclegeovsspectral}, a cycle does not form. In Figure \ref{fig:cycleCE2}(a), the system seems to reach the same low vorticity state as the nonlinear simulation during its transient, but then becomes stuck in a small region of energy space that has a higher potential energy than any of the points in the nonlinear elliptical cycle. In Figure \ref{fig:cycleCE2}(b), the system immediately finds itself stuck in this same region of high potential and kinetic energy, and performs a small cycle there. 

The inability to transition back to the high magnetic potential state in CE2 shows that these transitions are dependent on processes that cannot be captured by a quasilinear approximation. It may be that they owe their origin to eddy-eddy $\rightarrow$ eddy interactions, since these interactions are forbidden in CE2. The single transition that occurs may be the result of lingering initialization energy, likely aided by the fact that the high vorticity mode's power is concentrated in low zonal modes, while the low vorticity mode's power is more spread out, making a transition to the high vorticity mode possible without eddy-eddy interactions. We may hope that the problem is remedied by employing a GQL approximation.

\begin{figure}
\begin{center}
\includegraphics[width=14cm]{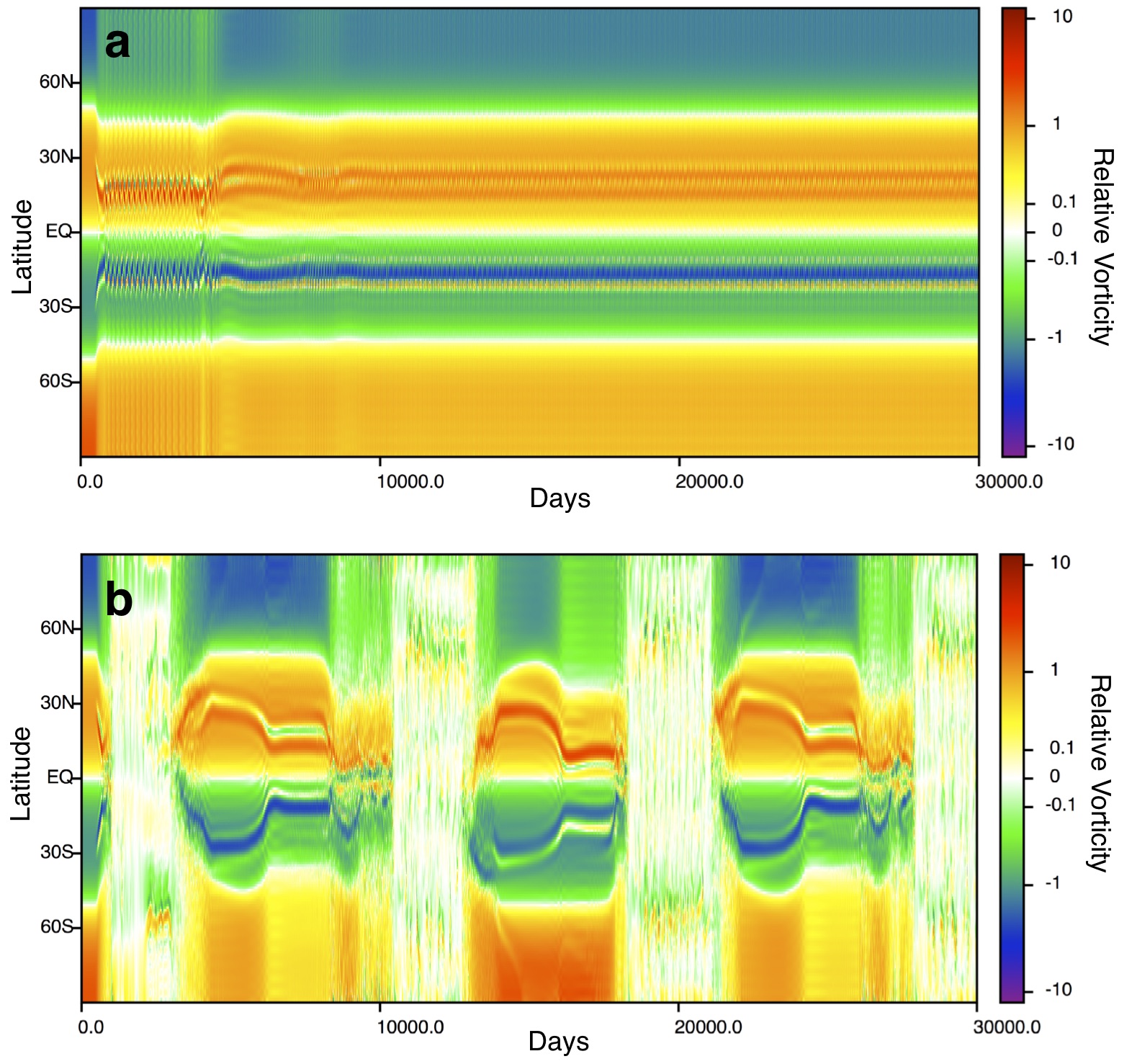}
\caption{{\label{fig:gql} GQL simulations at $R_m=2800$. (a): GQL simulation with $\Lambda=1$. (b): GQL simulation with $\Lambda=3$. Resolution remains $L_{max}=60$ and $M_{max}=15$.
}}
\end{center}
\end{figure}

\begin{figure}
\begin{center}
\includegraphics[width=14cm]{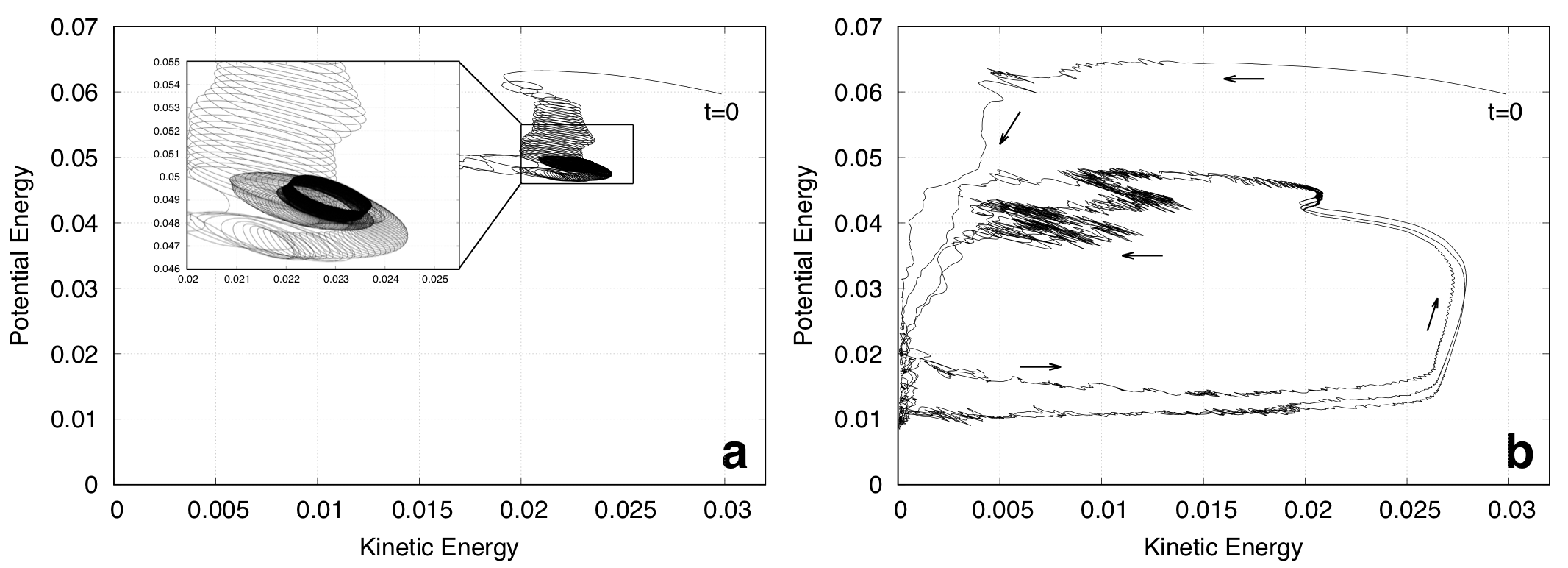}
\caption{{\label{fig:cycleGQL} GQL simulations in kinetic energy vs. potential energy space for $R_m$=2800. (a): GQL simulation with $\Lambda=1$, corresponding to Figure \ref{fig:gql}a. (b): GQL simulation with $\Lambda=3$, corresponding to Figure \ref{fig:gql}b. 
  }}
\end{center}
\end{figure}

The bifurcation diagram in Figure~\ref{fig:bifn} shows that GQL with $\Lambda=1$ or $\Lambda=3$ indeed performs better than CE2 in reproducing the NL results for all $R_m$.  
Figures~\ref{fig:gql} and \ref{fig:cycleGQL} show the relative vorticity timelines and the corresponding phase plane analysis for both of these thresholds $\Lambda$ at $R_m=2800$. These timelines illustrate that GQL with $\Lambda=1$ performs similarly to CE2 at this choice of parameter, yielding a weakly oscillatory state. However, increasing the threshold to $\Lambda=3$ allows the GQL approximation to reproduce the relaxation oscillation. Remarkably, GQL reproduces both the amplitude and period of the oscillation at this threshold, which is a testament to the effectiveness of this approximation.

\section{Conclusions}
\label{Conclusions}

We have shown that a simple, two-dimensional tachocline model can reproduce periodic transitions qualitatively similar to the seasonal exchange between toroidal field and wave-like dynamics observed in the solar record. The model undergoes complicated nonlinear transitions as the magnetic Reynolds number, $R_m$, is increased, with both hysteresis and relaxation oscillations emerging naturally. This near-heteroclininc behavior has, to the authors' knowledge, not yet been seen in joint instability simulations, though oscillations have also been seen in sustained magnetoshear instabilities \citep{Miesch_2007} and quasiperiodic energy exchange between toroidal field and magnetic Rossby waves have also recently been observed \citep{dikpati_etal_2017} both in models and in solar observations. Our simple model is less expensive than three dimensional thin-shell models or shallow water models that are often studied \citep{Miesch_2007, Gilman_2002}, which allows us to examine long timescale behavior at the cost of restricting instabilities to two dimensions.

We have also assessed the efficacy of the quasilinear (QL) and generalized quasilinear (GQL) approximations in reproducing these dynamics and that of direct statistical simulation (DSS at CE2) in reproducing the statistics. We find that CE2 is capable of reproducing many of the nonlinear states (and indeed the hysteresis between states), but for the most complicated relaxation oscillation states, CE2 cannot reproduce the transitions between states. This is another example of the  limitation of quasilinear models, that may become less relevant as the system moves away from statistical equilibrium \citep{tm2013}. We speculate that the addition of noise to DSS might enable this statistical formalism to reproduce the transitions. On the other hand, GQL appears to contain enough nonlinearity to represent the complicated nonlinear relaxation oscillation. This result is encouraging for a program of direct statistical simulation based on GQL or upon DSS formulations predicated on more sophisticated averaging procedures than simple zonal averaging \citep{Bakas:2013bo,Bakas:2014gf,Allawala:2017db}.

\acknowledgements {AP was supported in part by a Brown University LINK award. SMT was supported partially by a Leverhulme Fellowship and partially by funding from the European Research Council (ERC) under the European Union’s Horizon 2020 research and innovation programme (grant agreement no. D5S-DLV-786780) }

\bibliographystyle{jpp}
\bibliography{biblio}

\end{document}